\documentclass[aps,nofootinbib]{revtex4}

\usepackage{amsfonts,color}
\usepackage{graphicx,calc,epsfig}

\definecolor{blue}{rgb}{0,0,1}
\definecolor{green}{rgb}{0,1,0}
\definecolor{red}{rgb}{1,0,0}
\definecolor{van}{rgb}{1,0,1}
\definecolor{al}{rgb}{1,1,0}

\newcounter{mnotecount}[section]

\newcommand{\N}{\mathbb{N}}
\newcommand{\C}{\mathbb{C}}
\newcommand{\be}{\nopagebreak[3]\begin{equation}}
\newcommand{\ee}{\end{equation}}
\newcommand{\ba}{\nopagebreak[3]\begin{eqnarray}}
\newcommand{\ea}{\end{eqnarray}}
\DeclareFontFamily{U}{rsfs}{}         
\DeclareFontShape{U}{rsfs}{m}{n}{<5> rsfs5 <6><7> rsfs7          %
  <8><9><10><10.95><12><14.4><17.28><20.74><24.88> rsfs10}{}     %
\DeclareMathAlphabet{\mathfs}{U}{rsfs}{m}{n}                     %
\newcommand{\mfs}[1]{\mathfs {#1}}                               %
\newcommand{\va}{\scriptscriptstyle}

\newcommand{\Ss}{{\mfs S}}

\newcommand{\g}{\mathfrak{g}}

\newcommand{\SU}{\mathrm{SU}}
\newcommand{\SO}{\mathrm{SO}}
\newcommand{\so}{\mathfrak{so}}

\newcommand{\tr}{\mathrm{tr}}
\newcommand{\Aut}{\mathrm{Aut}}
\newcommand{\Hom}{\mathrm{Hom}}
\newcommand{\End}{\mathrm{End}}

\usepackage{psfrag}
\usepackage{vmargin}
\usepackage{amsfonts}
\usepackage{amsmath}
\usepackage{graphics,graphicx,color,calc,epsfig}
\usepackage{euscript}
\newcommand{\beq}{\begin{equation}}
\newcommand{\eeq}{\end{equation}}
\newcommand{\beqa}{\begin{eqnarray}}
\newcommand{\eeqa}{\end{eqnarray}}
\newcommand{\nn}{\nonumber}

\newcommand{\ve}{\mathcal{V}}

\newcommand{\R}{\mathbb{R}}
\newcommand{\V}{\mathbb{V}}
\newcommand{\T}{\Delta}

\newcommand{\h}{\hspace{1mm}}

\newcommand{\Si}{\Sigma}

\newcommand{\sW}{{\mfs W}}
\newcommand{\bh}{\mathcal{H}}
\newcommand{\M}{\Sigma_{\eta}}
\newcommand{\Reg}{\mathcal{R}_{(\eta,\epsilon)}}
\newcommand{\vphi}{\varphi}

\newtheorem{Pn}{Proposition}
\newtheorem{Lm}{Lemma}
\newtheorem{df}{Definition}
\newtheorem{Th}{Theorem}
\newtheorem{co}{Corollary}
\setpapersize{USletter}
\setmarginsrb{25mm}{12mm}{27mm}{1mm}{12pt}{11mm}{0pt}{13mm}

\begin{document}

\title{Extended matter coupled to BF theory}
\author{Winston J. Fairbairn \footnote{email: winston.fairbairn@ens-lyon.fr}}
\affiliation{Laboratoire de Physique - CNRS-UMR 5672 \\
ENS Lyon \\
46, All\'ee d'Italie \\
F-69364 Lyon, EU}
\author{Alejandro Perez \footnote{email: perez@cpt.univ-mrs.fr}}
\affiliation{Centre de Physique Th\'eorique
\footnote{Unit\'e mixte de recherche du CNRS et des Universit\'es Aix-Marseille I, Aix-Marseille II et du Sud Toulon-Var. Laboratoire affili\'e \`a la FRUNAM (FR 2291).} - CNRS-UMR 6207 \\
Luminy - Case 907 \\
F-13288 Marseille, EU}

\date{\today \vbox{\vskip 2em}}

\begin{abstract}
Recently, a topological field theory of membrane-matter coupled to BF theory in arbitrary spacetime dimensions was proposed \cite{BP}. In this paper, we discuss various aspects of the four-dimensional theory. Firstly, we study classical solutions leading to an interpretation of the theory in terms of strings propagating on a flat spacetime. We also show that the general classical solutions of the theory are in one-to-one correspondence with solutions of Einstein's equations in the presence of distributional matter (cosmic strings). Secondly, we quantize the theory and present, in particular, a prescription to regularize the physical inner product of the canonical theory. We show how the resulting transition amplitudes are dual to evaluations of Feynman diagrams coupled to three-dimensional quantum gravity. Finally, we remove the regulator by proving the topological invariance of the transition amplitudes.
\end{abstract}

\maketitle

\section{Introduction}

Based on the seminal results \cite{old} of $2+1$ gravity coupled to point sources, recent developments \cite{effqg}, \cite{karim} in the non-perturbative approach to $2+1$ quantum gravity have led to a clear understanding of quantum field theory on a three-dimensional quantum geometrical background spacetime. The idea is to first couple free point particles to the gravitational field {\itshape before} going through the second quantization process. In this approach, particles become local conical defects of spacetime curvature and their momenta are recasted as holonomies of the gravitational connection around their worldlines. It follows that momenta become group valued leading to an effective notion of non-commutative spacetime coordinates. The Feynman diagrams of such theories are related via a duality transformation to spinfoam models.

All though conceptually very deep, these results remain three-dimensional. The next step is to probe all possible extensions of these ideas to higher dimensions. Two ideas have recently been put forward. The first is to consider that fundamental matter is indeed pointlike and study the coupling of worldlines to gravity by using the Cartan geometric framework \cite{wise} of the McDowell-Mansouri formulation of gravity as a de-Sitter gauge theory \cite{mm}. The second is to generalize
the description of matter as topological defects of spacetime curvature to higher dimensions. This naturally leads to matter excitations supported by co-dimension two membranes \cite{BP}, \cite{BFloops}. Before studying the coupling of such sources to quantum gravity, one can consider, as a first step, the BF theory framework as an immediate generalization of the topological character of three-dimensional gravity to higher dimensions.

This paper is dedicated to the second approach, namely the coupling of string-like sources to BF theory in four dimensions.
The starting point is the action written in \cite{BP} generating a theory of flat connections except at the location of two-dimensional surfaces, where the curvature picks up a singularity, or in other words, where the gauge degrees of freedom become dynamical.
The goal of the paper is a two-fold. Firstly, acquire a physical intuition of the algebraic fields involved in the theory which generalize the position and momentum Poincar\'e coordinates of the particle in three-dimensions. Secondly, provide a complete background independent quantization of the theory in four dimensions, following the work done in \cite{BP}.

The organization of the paper is as follows. In section II, we study some classical solutions guided by the three-dimensional example. We show that some specific solutions lead to the interpretation of rigid strings propagating on a flat spacetime. More generally, we prove that the solutions of the theory are in one-to-one correspondence with distributional solutions of general relativity. In section III, we propose a prescription for computing the physical inner product of the theory. This leads us to an interesting duality between the obtained transition amplitudes and Feynman diagrams coupled to three-dimensional gravity.
We finally prove in section IV that the transition amplitudes only depend on the topology of the canonical manifold and of the spin network graphs.

\section{Classical theory}

\subsection{Action principle and classical symmetries}

Let $G$ be a Lie group with Lie algebra $\mathfrak{g}$ equipped
with an $Ad(G)$-invariant, non degenerate bilinear form noted
`$\tr$' (e.g. the Killing form if $G$ is semi-simple). Consider
the principal bundle $\mathcal{P}$ with $G$ as structure group and
as base manifold a $d+1$ dimensional, compact, connected, oriented
differential manifold $M$. We will assume that $\mathcal{P}$ is
trivial, all though it is not essential, and chose once and for
all a global trivialising section. We will be interested in the
following first order action principle, describing the interaction
between closed membrane-like sources and BF theory \cite{BP}: \beq
\label{totalaction} S[A,B;q,p]= S_{BF}[A,B] - \int_{\sW} \tr (B +
d_{A} q) p ). \eeq The action of free BF theory in $d+1$
dimensions is given by \beq \label{BFaction} S_{BF}[A,B] =
\frac{1}{\kappa} \int_{M} \tr( B \wedge F [A] ). \eeq Here, $B$ is
a $\mathfrak{g}$-valued $(d-1)$-form on $M$, $F$ is the curvature
of a $\mathfrak{g}$-valued one-form $A$, which is the pull-back to
$M$ by the global trivializing section of a connection on
$\mathcal{P}$, and $\kappa \in \R$ is a coupling constant.

In the coupling term, $\sW$ is the $(d-1)$-brane worldsheet defined by the embedding $\phi : E \subset \R^{d-1} \rightarrow M$,
$d_{A}$ is the covariant derivative with respect to the connection $A$, $q$ is a $\mathfrak{g}$-valued $(d-2)$-form on $\sW$ and $p$ is a $\mathfrak{g}$-valued function on $\sW$. The physical meaning of the matter variables $p$ and $q$ will be discussed in the following section. Essentially, $p$ is the momentum density of the brane and $q$ is the first integral of the $(d-1)$-volume element; the integral of a line and surface element in in three and four dimensions ($d=2,3$) respectively.

The equations of motion governing the dynamics of the theory are those of a topological field theory:
\ba
\label{em}
F[A]= \kappa p \ \delta_{\sW}
\label{flatness} \\
d_{A} B= \kappa [p, q] \delta_{\sW}
\label{impy} \\
\phi^* (B + d_A q) = 0
\label{q} \\
d_A p|_{\sW}=0.
\label{conservation}
\ea
Here, $\delta_{\sW}$ is a distributional two-form, also called current, which has support on the worldsheet $\sW$. It is defined such that for all $(d-1)$-form $\alpha$, $\int_{\sW} \alpha = \int_M (\alpha \wedge \delta_{\sW})$. The symbol $\phi^*$ denotes the pull-back of forms on $\sW$ by the embedding map $\phi$.

We can readily see that the above action describes a theory of local conical defects along brane-like $(d-1)$-submanifolds of $M$ through the first equation. The second states that the obstruction to the vanishing of the torsion is measured by the commutator of $p$ and $q$. The third equation is crucial. It relates the background field $B$ to the dynamics of the brane. For instance, this equation describes the motion of a particle's position in 3d gravity \cite{desousa}. The last states that the momentum density is covariantly conserved along the worlsheet. It is in fact a simple consequence of equation \eqref{flatness} together with the Bianchi identity $d_A F =0$. We will see how this is a sign of the reducibility of the constraints generated by the theory.

The total action is invariant under the following (pull back to $M$ of) vertical automorphisms of $\mathcal{P}$,
\beqa
\label{gravy1}
\forall g \in C^{\infty}(M, G ), \hspace{8mm}
B & \mapsto & B = g B g^{-1} \\ \nn
A & \mapsto & A = g A g^{-1} + g d g^{-1} \\ \nn
p & \mapsto & g p g^{-1} \\ \nn
q & \mapsto & g q g^{-1} \nn
\eeqa
and the `topological', or reducible transformations
\beqa
\label{gravy2}
\forall \eta \in \Omega^{d-2}(M,\mathfrak{g}), \hspace{8mm}
B & \mapsto & B + d_A \eta \\ \nn
A & \mapsto & A \\ \nn
p & \mapsto & p \\ \nn
q & \mapsto & q - \eta \nn
\eeqa
where $\Omega^{p}(M,\mathfrak{g})$ is the space of $\g$-valued $p$-forms on $M$.

\subsection{Physical interpretation:  the flat solution}

In this section, we discuss some particular solutions of the theory leading to an interpretation of matter propagating on flat backgrounds. We discuss the $d=2$ and $d=3$ cases where the gauge degrees of freedom of BF theory become dynamical along one dimensional worldlines and two-dimensional worldsheets respectively.

\subsubsection{The point particle in $2+1$ dimensions}

We now restrict our attention to the $d=2$ case with structure group the isometry group $G=\SO(\eta)$ of the diagonal form $\eta$ of a three-dimensional metric on $M$; $\eta=(\sigma^2,+,+)$ with $\sigma=\{1,i\}$ in respectively Riemannian ($G=\SO(3)$) and Lorentzian ($G=\SO(1,2)$) signatures. We denote $(\pi,V_{\eta})$ the vector (adjoint) representation of $\so(\eta) = \R\{J_{a}\}_{a=0,1,2}$, i.e., $V_{\eta} = \R^3$ and $V_{\eta} = \R^{1,2}$ in Riemannian and Lorentzian signatures respectively. The bilinear form `$\tr$' is defined such that $\tr(J_a J_b) = \frac{1}{2} \eta_{ab}$.
In this case, the free BF action \eqref{BFaction} describes the dynamics of three-dimensional general relativity, where the $B$ field plays the role of the triad $e$.
The matter excitations are $0$-branes, that is, particles and the worldsheet $\sW$ reduces to a one-dimensional worldline that we will note $\gamma$.
The degrees of freedom of the particle are encoded in the algebraic variables $q$ and $p$ which are both
$\so(\eta)$-valued functions with support on the world-line $\gamma$.

Firstly, we consider the open subset $U$ of $M$ constructed as
follows. Consider the three-ball $B^3$ centered on a point $x_0$
of the worldline $\gamma$ and call $x$ and $y$ the two punctures
$\partial B^3 \cap \gamma$. Pick two non intersecting paths
$\gamma_1$ and $\gamma_2$ on $\partial B^3$ both connecting $x$ to
$y$. The open region bounded by the portion of $\partial B^3$
contained between the two paths and the two arbitrary non
intersecting disks contained in $B^3$ and  bounded by the loops
$\gamma \gamma_1$ and $\gamma \gamma_2$ defines the open subset $U
\subset M$.

Next, we define the coordinate function $X: M \rightarrow
V_{\eta}$ mapping spacetime into the `internal space' $\so(\eta)$
isomorphic, as a vector space, to its vector representation space
$V_{\eta}$. The coordinates are chosen to be centered around a
point $x$ in $M$ traversed by the worldline; $X(x)=0$. Associated to
the coordinate function $X$, there is a natural solution to the equations of
motion \eqref{flatness}, \eqref{impy}, \eqref{conservation},
\eqref{q} in $U$ \ba e = d X = \delta \\ \nn A = 0 \\ \nn q = -X \mid_{\gamma}
\\ \nn p = constant, \ea
where $\delta$ is the unit of $\End (T_pM, V_{\eta})$,
$\delta(v)=v$ forall $v$ in $T_pM$ and all $p$ in $U$. The field configuration
$e=\delta$ (together with the $A=0$ solution) provides a natural
notion of flat Riemannian or Minkowskian spacetime geometry via
its relation to the spacetime metric $g= 2 \tr (e \otimes e)$. 
This flat background is defined in terms of a special gauge (notice 
that one can make $e$ equal to zero by transformation of the form 
\eqref{gravy2}). From now on, we will call such gauge a flat gauge.
The solution for $q$ is obtained through the equation \eqref{q}
relating the background geometry to the geometry of the worldline.
Here, we can readely see that $q$ represents the particle's
position $X$, first integral of the line element defined by the
background geometry $e$. Below we show that equation \eqref{impy}
forces the worldline to be a straight line. Finally, $p= constant$
trivially satisfies the conservation equation
\eqref{conservation}. In fact, the curvature equation of motion
\eqref{flatness} constrains $p$ to remain in a fixed adjoint orbit
so we can introduce a constant $m \in \R^{*+}$ such that $p = m v$
with $v\in \so(\eta)$ such that $\tr v^2 =-\sigma^2$.
Consequently, $p$ satisfies the mass shell constraints $p^2 := \tr
(p^2) = -\sigma^2 m^2$  and acquires the interpretation of the
particle's momentum.

We can now relate the position $q$ and momentum $p$, independent
in the first order formulation, by virtue of \eqref{impy}. Indeed,
the chosen flat geometry solution $e=\delta$, $A=0$ leads to a
everywhere vanishing torsion $d_A e$. Hence, the commutator $[p,q]
= X \times p$, where $\times$ denotes the usual cross product on
$V_{\eta}$, vanishes on the worldline. This vanishing of the
relativistic angular momentum (which is conserved by virtue of
equation \eqref{em}) implies, together with the flatness of the background
fields, that the the worldline $\gamma$ of the particle defines a straight line
passing through the origin and tangent to its momentum $p$.
Equivalently, we can think of the momentum $p$ as Hodge dual to a
bivector $*p$, in which case the worldline is normal to the plane
defined by $*p$.

Note that translating $\gamma$ off the origin, which requires the
introduction of spacetime torsion, can be achieved by the gauge
transformation $q\rightarrow q+C$ with $C=constant$ which leaves
all the other fields invariant. In this way we  conclude that the
previous solution of our theory can be (locally) interpreted as
the particle following a geodesic of flat spacetime.

More formally, we can also recover the action of a test particle
in flat spacetime by simply `switching off' the interaction of the
particle whith gravity. This can be achieved by evaluating the
action \eqref{totalaction} on the flat solution and neglecting the
interactions between geometry and matter, namely the equations of
motion linking the background fields to the matter degrees of
freedom (e.g. $e \neq dX$). This formal manipulation leads to the
following Hamilton function \be S[p,X,N]=\int_{\gamma} \tr( p \dot
X ) + N(p^2-m^2), \ee which is the standart first order action for a
relativistic spinless particle.

\subsubsection{The string in $3+1$ dimensions}

We now focus on the four dimensional ($d=3$) extension of the
above considerations. Here again we consider the isometry group
$G=\SO(\eta)$ of a given four dimensional metric structure
$\eta=(\sigma^2,+,+,+)$, in which case the value $\sigma=1$ leads
to the Riemannian group $G=\SO(4)$, while $\sigma=i$ encodes a
Lorentzian signature $G=\SO(1,3)$. As in three dimensions, we
denote $(\pi,V_{\eta})$, with $V_{\eta}=\R\{ e_I\}_I$, 
$I=0,...,3$, the vector representation of $\so(\eta) =
\R\{J_{ab}\}_{a,b=0,...,3}$.
Finally, we choose the bilinear form `$\tr$' such that, forall $a,b$ in $\so(\eta)$, it is
associated to the trace $\tr (ab) = \frac{1}{2} a_{IJ} b^{IJ}$ in
the vector representation. We are
using the notation $\alpha^{IJ}=\alpha^{ab} \,
\pi(J_{ab})^{IJ}:=\alpha^{ab} \, J_{ab}^{IJ}$ for the matrix
elements of the image of an element $\alpha \in \so(\eta)$ in
$\End(V_{\eta})$ under the vector representation.
The dynamics of the theory is governed by the action
\eqref{totalaction} where the matter excitations are string-like
and the worldsheet $\sW$ is now a two-dimensional submanifold of
the four dimensional space time manifold $M$. The string degrees
of freedom are described by an $\so(\eta)$-valued one-form $q$ and
an $\so(\eta)$-valued function $p$ living on the world-sheet
$\sW$.

As before, we construct an open subset $U \subset M$ by cutting
out a section of the four-ball $B^4$, and define the coordinate
function $X: M \rightarrow V_{\eta}$,  centered around a point $x$
in $M \cap \sW$. Consider the following field configurations which
define a flat solution to the equations of motion
\eqref{flatness}, \eqref{impy}, \eqref{conservation}, \eqref{q} in
$U$.: \ba B &=& * (e \wedge e), \;\; \mbox{with} \;\; e = dX =
\delta \\ \nn A &=& 0 \\ \nn q &=& - * X d X \\ \nn p &=&
constant, \ea where the star `$*$' is the Hodge operator
$*:\Omega^p(V_{\eta}) \rightarrow \Omega^{4-p}(V_{\eta})$ acting
on the internal space; $(* \alpha)_{IJ} = \frac{1}{2}
\epsilon_{IJ}^{\;\;\;KL} \alpha_{KL}$, with the totally
antisymmetric tensor $\epsilon$ normalized such that
$\epsilon^{0123}=+1$.

The solutions $B=*(\delta \wedge \delta)$ ($A=0$), leads to a
natural notion of flat Riemannian or Minkowski background geometry
through the standard construction of a metric out of $B$ when $B$
is a simple bivector; $B=*(e\wedge e)$ with $e=\delta$. We
can readily see that the $q$ one-form is the first integral of the
area element defined by the background field $B$. As in 3d, the equations
of motion constrain $p$ to remain in a fixed adjoint orbit so that
we can introduce a constant $\tau \in \R^{*+}$ such that  $p =
\tau v$ and $v\in \so(\eta)$ has a fixed norm; $\tr v^2 =
-\sigma^2$. We call $\tau$ the string tension, or mass per unit
length, and $p$ the momentum density which satisfies a generalized
mass shell constraint.

This momentum density $p$ is related to the $q$ field by analysis
of equation (\ref{impy}). The solution ($B=* \delta \wedge \delta,
A=0$) has zero torsion $d_A B$. Accordingly, the commutator
$[p,q]=[*X d X,p]$ vanishes on the worldsheet. This leads to the
constraint $X^{I}p_{IJ}=0$. Putting everything together, we see
that the flat solution in the open subset $U$ leads to the picture
of a locally flat worldsheet (a locally straight, rigid string) in
flat spacetime, dual \footnote{Note that this is exactly the same
result than the one obtained for the point particle, if we think
of the 3d momentum as Hodge dual to a bivector.}, as a
two-surface, to the momentum density bivector $p$ (if $p$ is simple, 
namely if it defines a two-plane). If we consider
more general solutions admitting torsion, the plane can be
translated off the origin. Indeed, the equation \eqref{q}
determines the field $q$ in terms of the geometry of the $B$ field
up to the addition of an exact one-form $\beta=d \alpha$ which
encodes the translational information. For instance, the
translation $X \mapsto X + C$ of the plane yields $q \mapsto q - *
C dX$ and consequently corresponds to a function $\alpha$ defined
by $d\alpha = - *C dX$. This potential is in turn determined by
the torsion $T=d_AB$ of the $B$ field via the equation
\eqref{impy}. More general solutions can be found for arbitrary
$\alpha$'s, as discussed below.

Following the same path as in the case of the particle case, we
can `turn off' the interaction between the topological BF
background and the string by evaluating the action on the flat
solution (this implies, here again, that we ignore the equations
of motion of the coupled theory, i.e. the relation between the
matter and geometrical degrees of freedom). We obtain the
following Hamilton function \be \label{effst}S[p,X,N]=\int_{\sW} \tr
(*p \; dX \wedge dX)+N(p^2-\tau^2), \ee up to a constant. This is
the Polyakov action on a non trivial background with metric
$G_{\mu\nu}=0$ and antisymmetric field $b=v\in \so(\eta)$.

Now, the previous action leads to trivial equations of motion that
are satisfied by arbitrary $X$ (because $p=constant$ and so the
Lagrangian is a total differential). This is to be expected, from
the string theory viewpoint, this would be a charged string moving
on a constant potential, so the field strength is zero. This seems
in sharp contrast to the particle case where the effective action
leads to straight line solutions. Here any string motion is
allowed; however, from the point of view of the full theory, all
these possibilities are pure gauge. The reason for this is that in
$2+1$ dimensions the flat gauge condition $e=\delta$ fixes the
freedom \eqref{gravy2} up to a global translation, and hence gauge
considerations are not necessary in  interpreting the effective
action. In the string case, $B=*(\delta \wedge \delta)$ partially
fixes the gauge; the remaining freedom being encoded in
$\eta=d\alpha$ for any $\alpha$.

\subsection{Geometrical interpretation: cosmic strings and topological defects}

The above discussion shows that particular solutions of the theory in a particular open subset lead to the standard propagation of matter degrees of freedom on a flat (or degenerate) background spacetime.
In fact, we can go further in the physical interpretation by considering other solutions, defined everywhere, which are in one-to-one correspondence with solutions of four dimensional general relativity in the presence of distributional matter. These solutions are called cosmic strings.

\subsubsection{Cosmic strings}

It is well known that the metric associated to a massive and
spinning particle coupled to three-dimensional gravity is that of
a locally flat spinning cone. The lift of this solution to $3+1$
dimensions corresponds to a spacetime around an infinitely thin
and long straight string (see for instance \cite{deser} and
references therein). Let us endow our spacetime manifold $M$ with
a Riemannian structure $(M,g)$ and let $x \in M$ label a point
traversed by the string. We can choose as a basis of the tangent
space $T_xM$ the coordinate basis $\{ \partial_t, \partial_r,
\partial_{\varphi}, \partial_z\}$ associated to local cylindrical
coordinates such that the string is lying along the $z$ axis and
goes through the origin. The embedding of the string is given by
$\phi(t,z)=(t,0,0,z)$. Let $\tau$ and $s$ respectively denote the
mass and intrinsic (spacetime) spin per unit length of the string.
Note that $\tau$ is the string tension. Solving Einstein's field
equations for a such stationary string carrying the above mass and
spin distribution produces a two-parameter $(\tau,s)$ family of
solutions described by the following line element written in the
specified cylindrical coordinates \beqa \label{cosmic} ds^2 &=&
\nonumber g_{\mu \nu} dx^{\mu} \otimes dx^{\nu} \\
     &=& \sigma^2 \left( dt + \beta d\varphi\right)^2 + dr^2 + (1-\alpha)^2 r^2 d\varphi^2 + dz^2 ,
\eeqa where $\beta = 4Gs$ and $\alpha = (1-4G\tau)$, $G$ is the
Newton constant. In fact this family of metrics is the general
solution to Einstein's equations describing a spacetime outside
any matter distribution in a bounded region of the plane
$(r,\varphi)$ and having a cylindrical symmetry. Exploiting the
absence of structure along the $z$ axis, by simply suppressing the
$z$ direction, reduces the theory to that of a point particle
coupled to gravity in $2+1$ dimensions, where the location of the
particle is given by the point where the string punctures the
$z=0$ plane. We will come across a such duality again in the
quantization process of the next sections. The dual co-frame for
the above metric is written \beqa \label{coframe} e^0 &=& dt +
\beta d \varphi \\ \nn e^1 &=& \cos \varphi d r - \alpha r \sin
\varphi d \varphi \\ \nn e^2 &=& \sin \varphi d r + \alpha r \cos
\varphi d \varphi \\ \nn e^3 &=& dz, \eeqa such that $ds^2=e^I
\otimes e^J \eta_{IJ}$.If we assume that the connection $A$
associated to the above metric is Riemannian, it is
straight-forward to calculate its components by exploiting
Cartan's first structure equation ($d_A e=0$). The result reads
\beq A = A^{IJ}_{\mu} \, \sigma_{IJ} d x^{\mu} = 4 G \tau \,
\sigma_{12} \, d \varphi , \eeq where $\{\sigma_{IJ}\}_{I,J}$
is a basis of $\Omega^2 ( V_{\eta} ) \simeq \so(\eta)$.

Using the distributional identity $d d \varphi = 2 \pi \delta^2(r)
dx dy$ ($x=r \cos \varphi$, $y=r \sin \varphi$, and $dx dy$ is a
wedge product), it is immediate to compute the torsion $T = T^0
e_0$ and curvature $F = F^{12} \, \sigma_{12}$ of the cosmic
string induced metric : \beq T^0 = 8 \pi G s \, \delta^2(r) \, dx
dy , \;\;\;\; F^{12} = 8 \pi G \tau \, \delta^2(r) \, dx dy . \eeq
These equations state that the torsion and curvature associated to
the cosmic string solution are zero everywhere except when the
radial coordinate $r$ vanishes, i.e. at the location of the string
worlsheet lying in the $z-t$ plane. If we now focus on the
spinless cosmic string case $s=0$, we can establish a one-to-one
correspondence between the above solutions of general relativity
and the following solutions of BF theory coupled to string
sources: \beq
\begin{array}{cccccc}
B^{01} &=& \sin \vphi dr dz + \alpha r \cos \vphi d \vphi dz, \;\;\;\; & B^{02} &=& -( \cos \vphi dz dr - \alpha r \sin \vphi dz d \vphi ) \\
B^{03} &=& \alpha r dr d \vphi, \;\;\;\; & B^{12} &=& - \sigma^2 dz dt \\
B^{13} &=& - \sigma^2 ( \sin \vphi dt dr + \alpha r \cos \vphi dt
d \vphi ), \;\;\;\; & B^{23} &=& \sigma^2 ( \cos \vphi dt dr -
\alpha r \sin \vphi dt d \vphi) , \nn
\end{array}
\eeq
\vspace{-3mm} \beqa && A^{12} = 4 G \tau d \varphi, \\ \nn &&
q^{12} = \sigma^2 (z dt - t dz), \;\;\;\; p^{12} = \tau,  \nn
\eeqa where only the non vanishing components have been written
and the coupling constant $\kappa$ in \eqref{totalaction} has been
set to $8 \pi G$.

In this way, solutions of our theory are in one-to-one
correspondence to solutions of Einstein's equations. The converse is
obviously not true as our model does not allow for physical local
excitations such us gravitational waves. However, augmenting the
action \eqref{totalaction} with a Plebanski term constraining the
$B$ field to be simple, would lead to the full Einstein equations
in the presence of distributional matter, \be \epsilon_{IJKL} e^J
\wedge F^{KL} = 8 \pi G \tau \, \epsilon_{IJKL} e^J J_{12}^{KL} \,
\delta_{\sW}, \ee where $J_{12}^{KL} = \delta^{[K}_1
\delta^{L]}_2$, starting from the theory considered in this paper.

\subsubsection{Many-strings-solution}

One can also construct a many string solution by `superimposing'
solutions of the previous kind at different locations. Here we
explicitly show this for two strings. We do this as the example
will illustrate the geometric meaning of torsion in our
model. Assume that we have two worlsheets $\sW_1$ and $\sW_2$
respectively traversing the points $p_1$ and $p_2$.
We will work with two open patches $U_i \subset M$, $i=1,2$, such that $p_1$ and $p_2$ both
belong to the overlap $U_1 \cap U_2$. The cylindrical coordinates
$(t_i,r_i,\varphi_i,z_i)$ associated to the charts $(U_i \subset
M, X^{\mu}_i : U_i \rightarrow \R^4)$ are chosen such that the
strings lie along the $z$ axis, are separated by a distance $x_0$
in the $x$-direction, and are such that $r_i(p_i)=0$. The
coordinate transform occurring in the overlap $U_1 \cap U_2$ is
immediate; it yields $t_i = t$, $x_2=x_1 + x_0$ $y_i=y$ and
$z_i=z$, for $i=1,2$. The two embeddings are given consequently by
$\phi_1(t,z)=(t,0,0,z)$ and $\phi_2(t,z)=(t,x_0,0,z)$. Our
notations are such that a field $\phi$ expressed in the coordinate
system associated to the open subset $U_i$ is noted $\phi_{U_i}$.

Our strategy to construct the two-string-solution is the
following. We need to realize the fact that, regarded from a
particular coordinate frame, one of the two strings is translated
off the origin. We will choose to
observe the translation of $\sW_2$ from the coordinate frame $1$.
Now, the study of the flat solution discussed in the previous
section has showed that translations of the worlsheet are related
to the torsion $T$ of the $B$ field. In particular, we know how to
recognize a translation of the form $X \rightarrow X + C$, with $C
= x_0 e_1$. It corresponds to a torsion of the form $T = \kappa
[p, d \alpha]$, with $d \alpha = -*C dX$. Hence, the
two-string-solution is based on the tetrad field which leads to
the desired value of the $B$ field torsion taking into account the
separation of the two worldsheets. For simplicity, here we assume
that the two strings are parallel, hence that they have same
momentum density \beq p_{U_1} =
p_{U_2} = \tau \sigma_{12}, \eeq and accordingly create the same
curvature singularity in both coordinate frames $1$ and $2$. The
associated connection yields \beq A_{U_i} = 4G \tau \, d \vphi_i
\, \sigma_{12}, \;\;\;\; \forall i = 1,2. \eeq The dual co-frame
$e_{U_i} = e_{U_i}^I \otimes e_{IU_i}$ is defined by the following
components \beqa \label{newcoframe}
e_{U_i}^0 &=& dt \\ \nn
e_{U_i}^1 &=& \cos \varphi_i d r_i - \alpha r_i \sin \varphi_i d \varphi_i \\ \nn
e_{U_i}^2 &=& \sin \varphi_i d r_i + (\alpha r_i \cos \varphi_i + \delta_{i2} \, \frac{\kappa}{4 \pi} \tau x_0) d \varphi_i \\ \nn
e_{U_i}^3 &=& dz . \eeqa By integrating the $B = * e \wedge e$
solution with $e$ given by \eqref{newcoframe}, we can now
calculate the $q$ field, up to the addition of an exact form
$\beta = d \alpha$ \beq q_{U_i} = \sigma^2 (z dt - t dz) \,
\sigma_{12} + d \alpha_i^{IJ} \sigma_{IJ}. \eeq The potential
$\alpha$ is derived from the equation of motion \eqref{impy}
relating the commutator of $p$ and $q$ to the $B$ field torsion
three-form $T = d_A B = * d_A e \wedge e + * e \wedge d_A e$ :
\beq T_{U_i} = \delta_{i2} \, \frac{1}{2} \, \kappa \, \tau x_0 \,
\delta(r) ( dx_2 \, dy_2 \, dz \, \sigma_{01} + \sigma^2 \, dt \,
dx_2 \, dy_2 \, \sigma_{13}). \eeq This torsion indeed corresponds
to a two-string-solution since it yields the desired value $-* C
dX$ for the form $d \alpha$, \beq d \alpha_i = \delta_{i2} \,
\frac{1}{2} \, x_0 \, ( dz \, \sigma_{02} + \sigma^2 \, dt \,
\sigma_{23} ). \eeq One can add more than one string in a similar
fashion, leading to multiple cosmic string solutions. It is
interesting to notice that torsion of the mutiple string solution
is related to the distance $x_0$ separating the world sheets. Of
course this is a distance defined in the flat-gauge where
$B=*\delta\wedge\delta$. This concludes our
discussion on the physical aspects of the action
\eqref{totalaction} of string-like sources coupled to BF theory.
We now turn toward the quantization of the theory.

\section{Quantum Theory}

For the entire quantization process to be well defined, we will restrict our attention to the case where the symmetry group $G$ is compact. For instance, we can think of $G$ as being $\SO(4)$. We will also concentrate on the four-dimensional theory and set the coupling constant $\kappa$ to one.
Also, to rely on the canonical analysis performed in \cite{BP}, we will work with a slightly different theory where the momentum $p$ is replaced by the string field $\lambda \in C^{\infty}(\sW,G)$. This new field enters the action only through the conjugation $\tau Ad_{\lambda}(v)$ of a fixed unit element $v$ in $\g$, and the theory is consequently defined by the action \eqref{totalaction} with $p$ set to $\tau \lambda v \lambda^{-1}$. The field $\lambda$ transforms as $\lambda \rightarrow g \lambda$ under gauge transformations of the type \eqref{gravy1} and the theory acquires a new invariance under the subgroup $H \subseteq G$ generated by $v$. The link between the two theories is established by the fact that, as remarked before, the equation of motion $F= p \delta_{\sW}$ implies that $p$ remains in the same conjugacy class along the worldsheet. Here, we choose to label the class by $\tau v$ and to consider $\lambda$ as dynamical field instead of $p$.

\subsection{Canonical setting}

As a preliminary step, we assume that the spacetime manifold $M$
is diffeomorphic to the canonical split $\R \times \Si$, where
$\R$ represents time and $\Si$ is the canonical spatial
hypersurface. The intersection of $\Si$ with the string worldsheet
$\sW$ forms a one dimensional manifold $\Ss$ that we will assume
to be closed\footnote{In fact if $\Si$ is compact the equations of
motion \eqref{flatness} implies that the string must be closed (or
have zero tension).}. We choose local
coordinates $(t,x^a)$ for which $\Sigma$ is given as the
hypersurface $\{t=0\}$. By definition, $x^a$, $a=1,2,3$, are local
coordinates on $\Sigma$. We also choose local coordinates $(t,s)$
on the $2$-dimensional world-sheet $\sW$, where $s \in [0,2\pi]$
is a coordinate along the one-dimensional string $\Ss$. We will
note $x_{\va \Ss} = \phi \mid_{\Sigma}$ the embedding of the
string $\Ss$ in $\Sigma$. We pick a basis
$\{X_i\}_{i=1,...,dim(\g)}$ of the real Lie algebra $\g$, raise
and lower indices with the inner product `$\tr$', and define
structure constants by $[X_i, X_j] = f_{ij}^{\;\;k} X_k$. Next, we
choose a polarization on the phase space such that the degrees of
freedom are encoded in the configuration variable $(A,\lambda) \in
\mathcal{A} \times \Lambda$ defined by the couples formed by (the
pull-back to $\Si$ of) connections and string momenta.

The canonical analysis of the coupled action \eqref{totalaction} shows that the legendre transform from configuration space to phase space is singular : the system is constrained. Essentially \footnote{See the original work \cite{BP} for a detailed canonical analysis.}, the constraints are first class and are given by the following set of equations :
\beqa
\label{gauss}
G_i &:=& D_{a}E^a_i + f_{ij}^{\;\;k} \, q^j_a \, \pi^a_k \, \delta_{\Ss} \, \approx \, 0 \\
\label{curvature1}
H_i^a &:=& \epsilon^{abc}F_{ibc} - \pi^a_i \, \delta_{\Ss} \, \approx \, 0.
\eeqa
Here, $E^a_i = \epsilon^{abc} B_{ibc}$ is the momentum canonically conjugate to $A_i^a$, $\pi^a_i=\partial_s \, x^a_{\va \Ss} p_i$ is conjugate to $q$ and satisfies $D_{a}\pi^a_i = 0$, where $p_i=\tr(X_ip)$ denote the components of the Lie algebra element $p$ in the chosen basis of $\g$. The symbols $D$ and $F$ denote respectively the covariant derivative and curvature of the spatial connection $A$.

The first constraint \eqref{gauss}, the Gauss law, generates kinematical gauge transformations while the second \eqref{curvature1}, the curvature constraint, contains the dynamical data of the theory. To quantize the theory, one can follow Dirac's program of quantization of constrained systems which consists in first quantizing the system before imposing the constraints at the quantum level. The idea is to construct an algebra $\mathfrak{A}$ of basic observables, that is, simple phase space functions which admit unambiguous quantum analogues, which is then represented unitarely, as an involutive and unital $\star$-algebra of abstract operators, on an unphysical or auxiliary Hilbert space $\mathcal{H}$.
Since the classical constraints are simple functionals of the basic observables, they can be unambiguously quantized, that is, promoted to self-adjoint operators on $\mathcal{H}$. The kernel of these constraint operators are spanned by the physical states of the theory.

The structure of the constraint algebra enables us to solve the constraints in different steps.
One can first solve the Gauss law to obtain a quantum kinematical setting. Then, impose the curvature constraint on the kinematical states to fully solve the dynamical sector of the theory.
In \cite{BP}, the kinematical subset of the constraints is solved and a kinematical Hilbert space $\mathcal{H}_{\mbox{\tiny{kin}}}$ solution to the quantum Gauss law is defined.
We first review the kinematical setting of \cite{BP} before exploring the dynamical sector of the theory.

\subsection{Quantum kinematics : the Gauss law}

The Hilbert space $\mathcal{H}_{\mbox{\tiny{kin}}}$ of solutions to the Gauss constraint is spanned by so-called string spin network states. String spin network states are the gauge invariant elements of the auxiliary Hilbert space $\mathcal{H}$ of cylindrical functions which is constructed as follows.

\subsubsection{Auxiliary Hilbert space $\mathcal{H}$}

Firstly, we define the canonical BF states. Let $\Gamma \subset
\Si$ denote an open graph, that is, a collection of one
dimensional oriented sub-manifolds \footnote{More precisely, one
usually endows the canonical hypersurface $\Si$ with a real,
analytic structure and restricts the edges to be piecewise
analytic or semi-analytic manifolds, as a mean to control the
intersection points.} of $\Si$ called edges $e_{\Gamma}$, meeting
if at all only on their endpoints called vertices $v_{\Gamma}$.
The vertices forming the boundary of a given edge $e_{\Gamma}$ are
called the source $s(e_{\Gamma})$ and target $t(e_{\Gamma})$
vertices depending on the orientation of the edge. We will call 
$n \equiv n_{\Gamma}$ the cardinality of the set of edges $\{e_{\Gamma}\}$ of
$\Gamma$. 
Let $\phi : G^{\times n} \rightarrow \C$ denote a continuous complex valued
function on $G^{\times n}$ and $A(e_{\Gamma}) \equiv
g_{e_{\Gamma}} =P \exp \, (\int_{e_{\Gamma}} A)$ denote the
holonomy of the connection $A$ along the edge $e_{\Gamma} \in
\Gamma$. The cylindrical function associated to the graph $\Gamma$
and to the function $\phi$ is a complex valued map
$\Psi_{\Gamma,\phi} : \mathcal{A} \rightarrow \C$ defined by :
\beq \Psi_{\Gamma,\phi}[A] =
\phi(A(e^1_{\Gamma}),...,A(e^n_{\Gamma})), \eeq forall $A$ in
$\mathcal{A}$. The space of such functions is an abelian
$\star$-algebra denoted Cyl$_{BF, \Gamma}$, where the
$\star$-structure is simply given by complex conjugation on $\C$.
The algebra of all cylindrical functions will be called Cyl$_{BF}$
= $\cup_{\Gamma}$ Cyl$_{BF, \Gamma}$.

Next, we define string states. Since the configuration variable is a zero-form, we expect to consider wave functions associated to points $x \in \Si$. Accordingly, we define the $\star$-algebra Cyl$_{S}$ of cylindrical functions on the space $\Lambda$ of $\lambda$ fields as follows. An element $\Phi_{X,f}$ of Cyl$_{S}$ is a continuous map $\Phi_{X,f} : \Lambda \rightarrow \C$, where $X = \{x_1,...,x_n\}$ is a finite set of points in $\Ss$ and $f : G^{\times n} \rightarrow \C$ is a complex valued function on the Cartesian product $G^{n}$, defined by
\beq
\Phi_{X,f} [\lambda] = f (\lambda(x_1),...,\lambda(x_n)).
\eeq
Both algebras Cyl$_{BF}$ and Cyl$_{S}$, regarded as vector spaces, can be given
a pre-Hilbert space structure. Fixing a graph $\Gamma \subset \Si$ with $n$ edges and a set of $m$ points $X \subset \Si$, we define the scalar products respectively on
Cyl$_{BF,\Gamma}$ and Cyl$_{S,p}$ as
\beq
\label{scalar1}
< \Psi'_{\Gamma,\phi} , \Phi_{\Gamma,\psi} > \, = \int_{G^{\times n}} \overline{\phi} \, \psi ,
\eeq
and
\beq
\label{scalar2}
< \Phi'_{X,f} , \Phi_{X,g} > \, = \int_{G^{\times m}} \overline{f} \, g ,
\eeq
where the integration over the group is realized through the Haar measure on $G$. These scalar products can be extended to the whole of Cyl$_{BF}$ (resp. Cyl$_{S}$), i.e. to cylindrical functions defined on different graphs (resp. set of points), by redefining a larger graph (resp. set of points) containing the two different ones. The resulting measure, precisely constructed via projective techniques, is the AL measure. The string Hilbert space was in fact introduced by Thiemann as a model for the coupling of Higgs fields to loop quantum gravity \cite{thomas} via point holonomies.
Completing these two pre-Hilbert spaces in the respective norms induced by the AL measures, one obtains the BF and string auxiliary Hilbert spaces respectively denoted $\mathcal{H}_{BF}$ and $\mathcal{H}_{S}$. Tensoring the two Hilbert spaces yields the auxiliary Hilbert space $\mathcal{H} = \mathcal{H}_{BF} \otimes \mathcal{H}_{S}$ of the coupled system.

Using the harmonic analysis on $G$, one can define an orthonormal basis in $\mathcal{H}_{BF}$ and $\mathcal{H}_{S}$ the elements of which are respectively denoted (open) spin networks and $n$-points spin states. Using the isomorphism of Hilbert spaces $L^2(G^{\times n}) \simeq \bigotimes_{e_{\Gamma}} L^2(G_{e_{\Gamma}})$, any cylindrical function $\Psi_{\Gamma,\phi}$ in $\mathcal{H}_{BF}$ decomposes according to the Peter-Weyl theorem into the basis of matrix elements of the unitary, irreducible representations of $G$ :
\beq
\Psi_{\Gamma,\phi}[A] = \sum_{\rho_1, ...,\rho_n} \phi_{\rho_1, ...,\rho_n} \, \rho_1 [A(e^1_{\Gamma})] \otimes ... \otimes \rho_n [A(e^n_{\Gamma})] ,
\eeq
where $\rho : G \rightarrow \Aut (\V_{\rho})$ denotes the unitary, irreducible representation of $G$ acting on the vector space $\V_{\rho}$ and the mode $\phi_{\rho_1, ...,\rho_n}:=\otimes_{i=1}^n \phi_{\rho_i}$ is an element of $( \h \V_{\rho_i} \otimes \V_{\rho_i} \,^* )^{\otimes_{i=1}^n}$. The functions appearing in the above sum are called open spin network states.

Equivalently, the string cylindrical functions decompose as :
\beq
\Phi_{X,f}[\lambda] = \sum_{\rho_1, ...,\rho_m} f_{\rho_1, ...,\rho_m} \, \rho_1 [\lambda(x_1)] \otimes ... \otimes \rho_m [\lambda(x_m)] ,
\eeq
and a given element in the sum is called an $n$-point spin state.

\subsubsection{String spin network states}

One can now compute a unitary action of the gauge group $C^{\infty}(\Si,G)$ on $\mathcal{H}$ by using the transformation properties of the holonomies and of the string fields $\lambda \rightarrow g \lambda$ under the gauge group and derive the subset of $G$-invariant states, that is, the states solution to the Gauss constraint. A vectorial basis of the vector space of gauge invariant states can be constructed, in analogy with $3d$ quantum gravity coupled to point particles \cite{karimale}, by tensoring the open spin network basis with the $n$ point spin states elements. Such an tensorial element is required the following consistency conditions to be $G$-invariant.

The graph $\Gamma$ of the open spin network has a set of vertices $\ve_{\Gamma}$ including the points $\{x_1,...,x_n\}$ forming the set $X$. The vertices of $\Gamma$ are coloured with a chosen element $\iota_v$ of an orthonormal basis of the vector space of intertwining operators
\beq
\Hom_G \h \left[ \bigotimes_{e_{\Gamma} \mid t(e_{\Gamma})=v_{\Gamma}} \mathbb{V}_{\rho_{e_{\Gamma}}} \h , \h \bigotimes_{e_{\Gamma} \mid s(e_{\Gamma})=v_{\Gamma}} \mathbb{V}_{\rho_{e_{\Gamma}}} \right],
\eeq
if the vertex $v_{\Gamma}$ is not on the string. If a vertex $v_{\Gamma}$ is on the string, it coincides with some point $x_k \in X$. In this case, we chose an element $\iota_{v_{\Gamma}}$ in an orthonormal basis of
\beq
\Hom_G \h \left[ \bigotimes_{e_{\Gamma} \mid t(e_{\Gamma})=v_{\Gamma}} \mathbb{V}_{\rho_{e_{\Gamma}}} \h , \h \left( \bigotimes_{e_{\Gamma} \mid s(e_{\Gamma})=v_{\Gamma}} \mathbb{V}_{\rho_{e_{\Gamma}}} \right) \otimes \mathbb{V}_{\rho_k} \right],
\eeq
where $\mathbb{V}_{\rho_k}$ is the representation space associated to the point $x_k$.

\begin{figure}[h]
\centerline{\hspace{0.5cm} \(
\begin{array}{c}
\includegraphics[height=6cm]{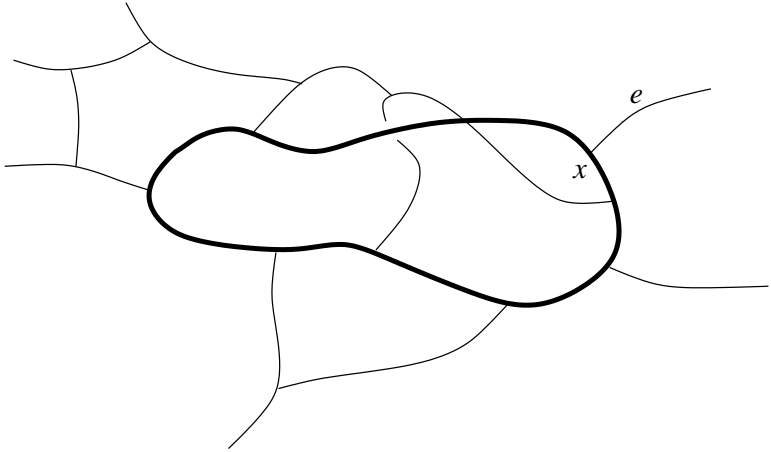}
\end{array}\)} \caption{A typical string spin network (the string is represented by the bold line).
}
\label{stringy}
\end{figure}

By finally implementing the invariance under the sub-group $H \subseteq G$ generated by $v$ of the $n$-point spin states by choosing the modes to be $H$-invariant, one obtains a vectorial basis in the kinematical Hilbert space $\mathcal{H}_{\mbox{\tiny{kin}}}$ where the inner product is that of (BF and string) cylindrical functions. The elements of this basis are called {\itshape string spin networks states} and are of the form (see fig. \ref{stringy}) :
\beq
\label{SSN}
\Psi_{\Gamma,X}[A,\lambda] := ( \Psi_{\Gamma} \otimes \Phi_X ) [A,\lambda]=  \left[ \bigotimes_{e_{\Gamma} \in \Gamma} \rho_{e_{\Gamma}}[A(e_{\Gamma})] \h \bigotimes_{x \in X} \rho_x[\lambda(x)] \right] \hspace{2mm}. \hspace{2mm} \bigotimes_{v_{\Gamma} \in \Gamma} \iota_{v_{\Gamma}} ,
\eeq
where the dot `$.$' denotes tensor index contraction.

This concludes the quantum kinematical framework of strings coupled to BF theory performed in \cite{BP}. We now solve the curvature constraint and compute the full physical Hilbert space $\mathcal{H}_{\mbox{\tiny{phys}}}$.

\subsection{Quantum dynamics: the curvature constraint}

In this section, we explore the dynamics of the theory by constructing the physical Hilbert space $\mathcal{H}_{\mbox{\tiny{phys}}}$ solution to the last constraint of the system, that is, the curvature or Hamiltonian constraint \eqref{curvature1}. Note that the physical states that we construct below are also solutions to the constraints of four dimensional quantum gravity coupled to distributional matter, as in the classical case.

We first underline a crucial property of the curvature constraint
of $d+1$-dimensional BF theory with $d>2$, namely its reducible
character which has to be taken into account during the
quantization process. We then proceed (as in \cite{karimale}) \`a
la Rovelli and Reisenberger \cite{rei,rovelli} by building and
regularizing a generalized projection operator mapping the
kinematical states into the kernel of the curvature constraint
operator. This procedure automatically provides the vector space
of solutions with a physical inner product and a Hilbert space
structure, and leads to an interesting duality with the coupling
of Feynman loops to 3d gravity \cite{barrett}, \cite{pr1} from the
covariant perspective.

\subsubsection{The reducibility of the curvature constraint}

A naive imposition of the curvature constraint on the kinematical
states leads to severe divergences. This is due to the fact that
there is a redundancy in the implementation of the constraint; the
components of the curvature constraint of $(d+1)$-dimensional BF
theory are not linearly independent, they are said to be
reducible, if $d>2$. The same is true for the theory coupled to
sources under study here. As an illustration of this fact, let us
simply count the degrees of freedom of source free ($\tau = 0$) BF
theory in $d+1$ dimensions.

The configuration variable of the theory $A_a^i$ is a $\g$-valued
connection one-form, thus containing $d \times \dim(\g)$
independent components for each space point of $\Si$. In turn, the
number of constraints is given by the $\dim(\g)$ components of the
Gauss law \eqref{gauss} plus the $d \times \dim(\g)$ components of
the Hamiltonian constraint \eqref{curvature1} for each space point
$x \in \Sigma$. Hence, we have $N_C= (d+1) \times \dim(\g)$
constraints per space point. This leads to a negative number of
degrees of freedom. What is happening \footnote{We thank Merced
Montesinos for pointing out this property of BF theory's
constraints.} ? The point is that the $N_C$ constraints are not
independent: the Bianchi identity ($D^{(2)} F= d^{(2)} F + [A,F] =
0$, where the superscript $^{(p)}$ indicates the degree of the
form acted upon) imply the reducibility equation \beq \label{redu}
D_a H_{i}^a = 0. \eeq In the case where sources are present, the
reducibility equation remains valid because the curvature
constraint $F=p\delta_{\Ss}$ together with the Bianchi identity
automatically implements the momentum density conservation $Dp=0$.
We will come back to this reducibility of the matter sector of the
theory. The system is said to be ($d-2$)-th stage reducible in the
first class curvature constraints. This designation is due to the
fact that the operator $d^{(2)}$ is himself reducible since
$d^{(3)} d^{(2)} \equiv 0$. In turn, $d^{(3)}$ is reducible and so
on. The chain stops after precisely $d-2$ steps since the action
of the $d^{(d)}$ de-Rham differential operator on $d$-forms is
trivial. Accordingly, the $N_R = \dim(\g)$ reducibility equations
\eqref{redu} imply a linear relation between the components of the
curvature constraint. The number $N_I$ of independent constraints
is thus given by $N_C-N_R = d \times \dim(\g)$. Using $N_I$ to
count the number of degrees of freedom leads to the correct
answer, namely zero degrees of freedom for topological BF theory.

The standard procedure to quantize systems with such reducible constraints consists in selecting a subset $H \mid_{irr}$ of constraints which are linearly independent and impose {\itshape solely} this subset of constraints on the auxiliary states of $\bh$.

Keeping this issue in mind, we now proceed to the definition and regularization of the generalized projector on the physical states and construct the Hilbert space $\mathcal{H}_{\mbox{\tiny{phys}}}$ of solutions to all of the constraints of the theory.

\subsubsection{Physical projector : formal definition - the particle/string duality}

We start by introducing the rigging map
\beqa
\eta_{\mbox{\tiny{phys}}} : \mbox{Cyl} &\rightarrow& \mbox{Cyl}^* \\ \nn
                              \Psi     &\mapsto& \delta (\hat{H}\mid_{irr}) \, \Psi,
\eeqa where Cyl$^*$ is the (algebraic) dual vector space of Cyl
$=$ Cyl$_{BF}$ $\otimes$ Cyl$_{S}$. The range of the rigging map
$\eta_{\mbox{\tiny{phys}}}$ formally lies in the kernel of the
Hamiltonian constraint of the coupled model. The power of the
rigging map technology is that it automatically provides the
vector space
$\eta_{\mbox{\tiny{phys}}}(\mbox{Cyl})=\mbox{Cyl}^*_{\mbox{\tiny{phys}}}
\subset$ Cyl$^*$ of solutions to the Hamiltonian constraint with a
pre-Hilbert space structure encoded in the physical inner product
\beq \label{physical} < \eta_{\mbox{\tiny{phys}}}(\Psi_1) ,
\eta_{\mbox{\tiny{phys}}}(\Psi_2) >_{\mbox{\tiny{phys}}} =
[\eta_{\mbox{\tiny{phys}}} (\Psi_2)](\Psi_1) := < \Psi_1 , \,
\delta(\hat{H}\mid_{irr}) \, \Psi_2 >, \eeq for any two string
spin network states $\Psi_1$, $\Psi_2$ $\in
\mathcal{H}_{\mbox{\tiny{kin}}}$. The scalar product used in the
last equality is the kinematical inner product \eqref{scalar1},
\eqref{scalar2}. The physical Hilbert space
$\mathcal{H}_{\mbox{\tiny{phys}}}$ is then obtained by the
associated Cauchy completion of the quotient of
$\mathcal{H}_{\mbox{\tiny{kin}}}$ by the Gel'fand ideal defined by
the set of zero norm states.

Accordingly, the construction of the physical inner product can
explicitly be achieved if we can rigorously make sense of the
formal expression $\delta(\hat{H} \mid_{irr})$. This task is greatly
simplified by virtue of the following duality. Indeed, we can re
express the above formal quantity as follows \beq
\label{projection} \delta (\hat{H}\mid_{irr}) = \prod_{x \in \Si}
\delta (\hat{H}\mid_{irr}(x)) = \int_{\mathcal{N}} \mathcal{D} \mu
[N] \h \exp \left( i \int_{\Si} \tr ( N \wedge \hat{H} ) \right).
\eeq Here, $\mathcal{N} \ni N$ is the space of regular
$\mathfrak{g}$-valued one-forms on $\Si$ and $\mathcal{D} \mu [N]$
denotes a formal functional measure on $\mathcal{N}$ imposing
constraints on the test one-form $N$ to remove the redundant delta
functions on $H$. Simply plugging in the explicit expression of
the exponent in \eqref{projection} leads to \beqa
\label{integralexponential}\nonumber
H[N]=\int_{\Si} \tr ( N \wedge H ) &=& \int_{\Si} \tr ( N \wedge F ) + \int_{\Ss} \tr ( N p ) \\
&=& S_{BF + part}^{3d}[N,A],
\eeqa
which, in the case where $G=SO(\eta)$, where $\eta$ is a three-dimensional metric, is the action of 3d gravity coupled to a (spinless) point particle \cite{desousa}, \cite{pr1} : the role of the triad is played by $N$, the mass and the worldline of the particle are respectively given by the string tension $\tau$ (hidden in the string variable $p = \tau Ad_{\lambda}(v)$) and $\Ss$. Finally, the role of the Cartan subalgebra generator $J_0$ is played by $v$ (also hidden in $p$). This relation is reminiscent to the link between cosmic strings in 4d and point particles in three-dimensional gravity discussed in the first sections. More generally, we have in fact the following duality :
\beq
\label{duality}
P(\Omega)_{BF + (d-2)-branes}^{d+1} \h = \h \mathcal{Z}_{BF + (d-3)-branes}^{d},
\eeq
where $\Omega$ denotes the ($d+1$-dimensional) no-spin-network vacuum state, $\mathcal{Z}$ is the path integral of BF theory in $d$ spacetime dimensions and we have introduced the linear form \footnote{More precisely, the linear form $P$, once normalized by the evaluation $P(1)$, is a state
$$P / P(1) : \mbox{Cyl} \subset \mathfrak{A} \rightarrow \C,$$
whose associated GNS construction leads equivalently to the physical Hilbert space $\mathcal{H}_{\mbox{\tiny{phys}}}$. The associated Gel'fand ideal $\mathcal{I}$ is immense by virtue of the topological nature of the theory under consideration. Indeed, one can show that any element of Cyl based on a contractible graph is equivalent to a complex number. The associated physical representation $\pi_{\mbox{\tiny{phys}}} : \mbox{Cyl} \rightarrow \End(\mathcal{H}_{\mbox{\tiny{phys}}})$ is defined such that forall cylindrical function $a_{\Gamma} \in \mbox{Cyl}$ defined on a contractible graph $\Gamma$, $\pi_{\mbox{\tiny{phys}}}(a_{\Gamma}[A]) \Psi = a_{\gamma}[0] \Psi$, forall $\Psi$ in $\mathcal{H}_{\mbox{\tiny{phys}}}$.} $P$ on Cyl $\subset \mathfrak{A}$ defined by
\beqa
\label{P}
\forall \Psi \in \mbox{Cyl}, \hspace{5mm} P(\Psi) &=& < \eta_{\mbox{\tiny{phys}}}(\Omega),  \eta_{\mbox{\tiny{phys}}}(\Psi) >_{\mbox{\tiny{phys}}} \\ \nn &=& < \Omega \,\, , \delta(\hat{H} \mid_{irr}) \,\, \Psi >.
\eeqa
Furthermore, when $d=3$, the formal functional measure $\mathcal{D} \mu [N]$ introduced above to take into account the reducible character of the four-dimensional theory corresponds to the Fadeev-Popov determinant gauge-fixing
the translational topological symmetry of the 3d theory;
the reducibility of the 4d theory is mapped via this duality onto the gauge redundancies of the three-dimensional theory.

Now because of the above duality \eqref{duality}, regularizing the
formal expression \eqref{projection} is, roughly speaking,
equivalent to regularizing the path integral for 3d gravity
coupled to point particles, up to the insertion of spin network
observables. The physical inner product in our theory will
therefore be related to amplitudes computed in \cite{pr1},
\cite{daniele}, although they would have here a quite different
physical interpretation. Following \cite{karimale, karimale2}, we
will regularize the Hamiltonian constraint at the classical level
by defining a lattice-like discretization of $\Si$ and by
constructing holonomies around the elementary plaquettes of the
discretization as a first order approximation of the curvature.

However, there are two major obstacles to the direct and naive
implementation of such a program. The first is the reducible
character of the curvature constraint and the second is the
presence of spin network edges ending on the string. We will use
the above duality to treat the first issue while the second will
be dealt with by introducing an appropriate regularization scheme.

\subsubsection{Physical projector : regularization}

Throughout this section, we will concentrate on the definition of
the linear form \eqref{P} evaluated on the most general string
spin network state $\Psi \in \mathcal{H}_{\mbox{\tiny{kin}}}$,
since it contains all the necessary information to compute
transition amplitudes between any two arbitrary elements of the
kinematical Hilbert space. We will consider a string spin network
basis elements $\Psi$ of $\mathcal{H}_{\mbox{\tiny{kin}}}$ defined
on the (open) graph $\Gamma$. The set of end points of the graph
living on the string $\Ss$ will be denoted $X$.

We follow the natural generalization of the regularization defined
in \cite{karimale} for $2+1$ gravity coupled to point particles.
In order to deal with the curvature singularity at the string
location, we thicken the smooth curve $\Ss$ to a torus topology,
smooth, non-intersecting tube $T_{\eta}$ of constant radius $\eta
> 0$ centered on the string $\Ss$. The radius $\eta$ is defined in terms
of the local arbitrary coordinate system.
If the string is disconnected,
we blow up each string component in a similar fashion.

Next, we remove the tube $T_{\eta}$ from the spatial manifold
$\Si$. We are left with a three-manifold with torus boundary
$\Sigma \setminus T_{\eta}$ noted $\M$. For instance, if $\Sigma$
has the topology of $S^3$, we know by Heegard's splitting, that the
resulting manifold has the topology of a solid torus whose boundary surface is the
Heegard surface defined by the string tube.
In this way we construct a new three-manifold with boundary where each boundary 
component is in one-to-one correspondence with a string component and has the
topology of a torus.
Finally, the open graph $\Gamma$ is embedded in the bulk
manifold and its endpoints lie on the boundary torus.

The next step is to choose a simplicial decomposition
\footnote{Note that in dimension $d \leq 3$, each topological
$d$-manifold admits a piecewise-linear-structure (this is the
so-called `triangulation conjecture').} of $\M$ or more generally
any cellular decomposition, i.e., a homeomorphism $\phi : \M
\rightarrow \T$ from our spatial bulk manifold $\M$ to a cellular
complex $\T$. The discretized manifold $\T \equiv \T_{\epsilon}$
depends on a parameter $\epsilon \in \R^{+}$ controlling the
characteristic (coordinate) `length scale' of the cellular
complex. We will see that, by virtue of the three-dimensional
equivalence between smooth, topological and piecewise-linear (PL)
categories, together with the background independent nature of our
theory, no physical quantities will depend on this extra
parameter. We will note $\Delta_k$ the $k$-cells of $\T$. To make
contact with the literature, we will in fact work with the dual
cellular decomposition $\T^*$. The dual cellular complex $\T^*$ is
obtained from $\T$ by placing a vertex $v$ in the center of each
three-cell $\Delta_3$, linking adjacent vertices with edges $e$
topologically dual to the two-cells $\Delta_2$ of $\T$, and
defining the dual faces $f$, punctured by the one-cells
$\Delta_1$, as closed sequences of dual edges $e$. The
intersection between $\T^*$ and the boundary tube $T_{\eta}$
induces a closed, oriented (trivalent if $\T$ is simplicial) graph
which is the one-skeleton of the cellular complex $\partial
\T^*=(\overline{v}, \overline{e}, \overline{f})$ dual to the
cellular decomposition $\partial \T$ of the 2d boundary $T_{\eta}$
induced by the bulk complex $\T$. We will note $\mathcal{F}$ the
set of faces $f$ of the cellular pair $(\T^*, \partial \T^*)$ and
require that each dual face of $\mathcal{F}$ admits an orientation
(induced by the orientation of $\M$) and a distinguished vertex.

Finally, among all possible cellular decompositions, we select a subsector of two-complexes which are {\itshape adapted} to the graph $\Gamma$. Namely, we consider dual cellular complexes $(\T^*, \partial \T^*)$ whose one-skeletons admit the graph $\Gamma$ as a subcomplex. In particular, the open edges of $\Gamma$ end on the vertices $\overline{v}$ of the boundary two-complex $\partial \T^*$.

The meaning of the curvature constraint $F = p \, \delta_{\Ss}$ is that the physical states have support on the space of connections which are flat everywhere except at the location of the string where they are singular. In other words,
the holonomy $g_{\gamma} = A(\gamma)$ of an infinitesimal loop $\gamma$ circling an empty, simply connected region yields the identity, while the holonomy $g_{\gamma}$ circling the string around a point $x \in \Ss$ is equal to $\exp p(x)$, the image of the fixed group element $u=e^{\tau v}$ under the inner automorphism $Ad_{\lambda} : G \rightarrow G$; $u \mapsto \lambda(x) \, u \, \lambda^{-1}(x)$, with the string field $\lambda$ evaluated at the point $x$. The integration over the string field $\lambda$ appearing in the computation of the physical inner product then forces the holonomy of the connection around the string to lie in the same conjugacy class $Cl(u)$ than the group element $u$.

To impose the $F=0$ part of the curvature constraint, we will require that the holonomy
\beqa
A : \mathcal{F} &\rightarrow& G \\ \nn
     \partial f &\mapsto& g_{f} = \prod_{e \subset \partial f} A(e),
\eeqa around all the oriented boundaries of the faces $f$ of
$\mathcal{F}$ be equal to one \footnote{Note that the blow up of
the string, reflected here in the presence of a flatness
constraint on the boundary torus, gives us the opportunity to
impose that the connection is flat also on the string.}. Each such
flat connection defines a monodromy representation of the
fundamental group $\pi_1(\M)$ in $G$. Concretely, the holonomies
are computed by taking the edges in the boundary $\partial f$ of
the face $f$ in cyclic order, following the chosen orientation,
starting from the distinguished vertex. Reversing the orientation
maps the associated group element to its inverse.

It is here crucial to take into account the reducible character of
the curvature constraint to avoid divergences due to redundancies
in the implementation of the constraints (i.e. divergences coming
from the incorrect product of redundant delta functions). As
discussed above, the reducibility equation induced by the Bianchi
identity implies that the components of the curvature are not
independent. In the discretized framework, we know \cite{LCS} that
forall set of faces $f$ forming a closed surface $\mathcal{S}$
with the topology of a two-sphere, \beq \label{bianchi} \prod_{f
\in \mathcal{S}} g_{f} = 1\!\!1 , \eeq modulo orientation and some
possible conjugations depending on the base points of the
holonomies. Accordingly, there is, for each three-cell of the dual
cellular complex $\T^*$, one group element $g_{f}$, among the
finite number of group variables attached to the faces bounding
the bubble, which is completely determined by the others. It
follows that imposing $g_f=1\!\!1$ on {\itshape all} faces of the
cellular complex $\T^*$ is redundant and would create divergences
in the computation of the physical inner product. The proper way
\cite{etera}, \cite{pr1} to address the reducibility issue, or
over determination of the holonomy variables, is to pick a maximal
tree $T$ of the cellular decomposition $\T$ and impose $F=0$
{\itshape only} on the  faces of $\T^*$ that are not dual to any
one simplex contained in $T$. A tree $T$ of a cellular
decomposition $\T$ is a sub-complex of the one-skeleton of $\T$
which never closes to form a loop. A tree $T$ of $\T$ is said to
be maximal if it is connected and goes through all vertices of
$\T$. The fact that $T$ is a maximal tree implies that one is only removing
redundant flatness constraints taking consistently into account
the reducibility of the flatness constraints.

Finally, we need to impose the $F=p$ part of the curvature
constraint. The idea is to require that the holonomy $g_{\gamma}$
around any loop $\gamma$ in $\partial \T^*$ based at a point $x$,
belonging to the homology class of loops of the boundary torus
$T_{\eta}$ normal to the string $\Ss$ (these loops are the ones
wrapping around the cycle of the torus circling the string, i.e.,
the non-contractible loops in $\M$), be equal to the image of the
group element $u$ under the adjoint automorphism $\lambda(x) u
\lambda(x)^{-1}$, i.e., belong to $Cl(u)$. Intuitively, this could
be achieved by picking a finite set $\{\gamma_i\}_i$ of such
homologous paths all along the tube $T_{\eta}$ and imposing
$g_{\gamma_i} = Ad_{\lambda_i} (u)$, with the field $\lambda_i$
evaluated at the base point of the holonomy $g_{\gamma_i}$.
However, here again, care must be taken in addressing the
reducibility issue induced by the equation $D_a H^a_i=0$. In the
presence of matter, the reducibility implies that the curvature
constraint $F=p\delta_{\Ss}$ together with the Bianchi identity
$DF=0$ induce the momentum density conservation $Dp=0$. In our
setting, this is reflected in the fact that the holonomies
$g_{\gamma_1}$ and $g_{\gamma_2}$ associated to two distinct
homologous loops $\gamma_1$ an $\gamma_2$ circling the string
satisfy the property $Cl(g_{\gamma_1})=Cl(g_{\gamma_2})$ on shell.
This is due to the Bianchi identity in the interior of the
cylindrically shaped section of the torus $T_{\eta}$ bounded by
$\gamma_1$ and $\gamma_2$, and the flatness constraint $F=0$
imposing the holonomies around all the dual faces on the boundary
of the cylindrical section to be trivial (see e.g.
\eqref{bianchi}). Accordingly, imposing $g_{\gamma_1} \in Cl(u)$
naturally implies that $g_{\gamma_2}$ belongs to the conjugacy
class labeled by $u$. In other words, choosing {\itshape one}
arbitrary closed path circling the string, say $\gamma_1$ based at
a point $x_1$, and imposing $F=p$ {\itshape only along that path}
naturally propagates via the Bianchi identity and the flatness
constraint and forces the holonomy $g_{\gamma_2}$ around any other
homologous loop $\gamma_2$ based at a point $x_2$ to be of the
form $g_{\gamma_2}=h u h^{-1} \, \in Cl(u)$, for some $h \in G$.
This shows that imposing $F=p$ more than once, e.g. also around
$\gamma_2$, would lead to divergences which can be traced back to
the reducibility of the constraints.

However, for the prescription to be complete \footnote{ To
understand these last points, consider two loops $\gamma_1$ and
$\gamma_2$ belonging to the same homology class circling the
string at two neighboring points, say $x_1$ and $x_2$. These two
loops define a section of the torus $T_{\eta}$ homeomorphic to a
cylinder. Suppose that the dual cellular complex $\T^*$ is such
that the cylinder is discretized by a single face with two
opposite edges glued along an dual edge $\beta$ in
$\partial \T^*$ connecting $x_1$ and $x_2$. The flatness
constraint on the boundary of the cylinder implies the following
presentation of the cylinder's fundamental polygon:
$$g_{\gamma_1} g_{\beta} g_{\gamma_2}^{-1} g_{\beta}^{-1}=1\!\!1,$$
which relates the holonomies $g_{\gamma_1}$ and $g_{\gamma_2}$ by virtue of the Bianchi identity in the interior of the tube. Hence, imposing $F=p$ only along one of the two loops, say $\gamma_1$, naturally leads to the constraint $g_{\gamma_2} \in Cl(u)$. Finally, plugging the relation $g_{\beta} = \lambda_{1} \lambda_{2}^{-1}$ in the value of the holonomy $g_{\gamma_2}$
leads to the required constraint $g_{\gamma_2} = \lambda_2 u \lambda_2^{-1}$.},
it is not sufficient to have $g_{\gamma_2}$ in $Cl(u)$;
we need to recover the fact that the holonomy along the loop $\gamma_2$ is the conjugation of the group element $u$ under the {\itshape dynamical field} $\lambda$ evaluated at the base point $x_2$ of the holonomy, namely $\lambda_2=\lambda(x_2)$.
This suggests an identification of the group element $h$ conjugating $u$ with the value of the string field $\lambda_2$, which leads to a relation between the holonomy $g_{\beta}$ along a path $\beta$ connecting the points $x_1$ and $x_2$ and the value of the string field $\lambda$ at $x_1$ and $x_2$:
$$g_{\beta} = \lambda_{s(\beta)} \lambda_{t(\beta)}\hspace{.01mm}^{-1},$$
stating that $\lambda$ is covariantly constant along the string. We have seen that the Bianchi identity together with the full curvature constraint induces the momentum density conservation. Our treatment of the reducibility issue consists in truncating the curvature constraint, i.e., in imposing $F=p$ only once, and using the Bianchi identity supplemented with the {\itshape momentum conservation} $D p=0$ to recover the truncated components of the curvature constraint without any loss of information.

Accordingly, the full prescription is defined via a choice of a closed, oriented path $\alpha$ and a finite set $C$ of open, oriented paths $\beta$ in $\partial \T^*$. The closed path $\alpha$ circles the string (it is non-contractible in the three-manifold $\M$). This loop is based at a point $x \in X$ lying on a dual vertex $\overline{v} \in \partial \T^*$ supporting a spin network endpoint. The open paths $\beta \in C$ are defined as follows. Let $\overline{\gamma} \in \partial \T^*$ be an oriented loop based at $x$, non-homologous to $\alpha$ (along the cycle of $T_{\eta}$ contractible in $\M$) and connecting all the spin network endpoints $x_k \in X$. Define the open path $\gamma$ by erasing the segment of $\overline{\gamma}$ supported by the edge $\bar{e}$ which is such that $x=t(\bar{e})$. The paths $\beta$ of $C$ are 1d sub-manifolds of $\gamma$ each connecting $x$ to a vertex $\bar{v}$ traversed by $\gamma$. If the graph $\Gamma$ is closed, one reiterates the same prescription simply dropping the requirements on the spin network endpoints $x_k \in X$, in particular, the base point $x$ is chosen arbitrarily.
We then impose $g_{\alpha} = \exp p$ with $p$ evaluated at the point $x$, and $g_{\beta} = \lambda_{s(\beta)} \lambda_{t(\beta)}\hspace{.01mm}^{-1}$, where $x=s(\beta)$, on each open path $\beta$ of $C$.

To summarize, we choose a regulator $\mathcal{R}_{(\eta,\epsilon)}
= (T_{\eta}, (\T_{\epsilon},\partial \T_{\epsilon}), T, \alpha,
C)$ consisting in a thickening $T_{\eta}$ of the string, a
cellular decomposition $(\T_{\epsilon},\partial \T_{\epsilon})$ of the manifold
$(\M,T_{\eta})$ adapted to the graph $\Gamma$, a maximal tree $T$
of $\T$, a closed path $\alpha$ in $\partial \T^*$, and a
collection $C$ of open paths $\beta$ in $\partial \T^*$. The
associated regularized physical scalar product is then given by
\beq \label{regulfinal1} P[\Psi] : =  \lim_{\eta, \epsilon
\rightarrow 0} P[\mathcal{R}_{(\eta,\epsilon)};\Psi], \eeq with
\beq \label{regulfinal2} P[\mathcal{R}_{(\eta,\epsilon)};\Psi] = <
\h \Omega \, , \left[ \prod_{f \notin T} \delta \left(g_{f}
\right) \; \prod_{\alpha} \delta \left(g_{\alpha} \exp p \right) \; \prod_{\beta
\, \in \, C} \delta(g_{\beta} \, \lambda_{t(\beta)} \,
\lambda_{s(\beta)}\hspace{.01mm}^{-1}) \right] \, \Psi \h > , \eeq
where the product over $\alpha$ is to take into account the possible 
multiple connected components of the string.

It is important to point out that, in addition to the expression of
the generalized projection above, we can use the regularization to
give an explicit expression of the regularized constraint
corresponding to $H[N]$ in equation (\ref{integralexponential}).
With the notation introduced so far the regulated quantum curvature constraint
becomes \be \hat{H}_{\eta, \epsilon}[N]=\sum_{f\in \Delta^*}{\rm
Tr}[N(x_f) g_f]+\sum_{\alpha} {\rm Tr}[N(x_{p}) g_{\alpha} \exp
p],\ee where $x_f$ is an arbitrary point in the interior of the
face $f$ and $x_p$ and arbitrary point on the string dual to the
loop $\alpha$ (the sum over $\alpha$ is over all the string
components). It is easy to check that the regulated quantum curvature
constraints satisfy off-shell anomaly freeness condition. For
instance \be U[g]\hat{H}_{\eta, \epsilon}[N]U^{\dagger}[g]=\hat{H}_{\eta,
\epsilon}[gNg^{-1}],\ee where $U[g]$ is the unitary generator of
$G$-gauge transformations. Therefore the regulator does not break
the algebraic structure of the classical constraints. The
quantization is consistent. The quantum constraint operator is
defined as the limit where $\eta$ and $\epsilon$ are taking to
zero. Instead of doing this in detail we shall simply concentrate
on the regulator independence of the physical inner product in the
following section.

The inner product is computed using the AL measure,
that is, by Haar integration along all edges of the graph $(\T^*,
\partial \T^*)$ $\ni \Gamma$ and along all endpoints $x_k \in X
\subset
\partial \T^*$ of the graph $\Gamma$.

We can now promote the classical delta function on the lattice phase space to a multiplication operator on $\mathcal{H}_{\mbox{\tiny{kin}}}$ by using its expansion in irreducible unitary representations :
\beq
\forall g \in G, \h \h \delta (g) = \sum_{\rho} \dim (\rho) \chi_{\rho} (g),
\eeq
where $\chi_{\rho} \equiv \tr \h \rho: G \rightarrow \C$ is the character of the representation $\rho$.
Each $\chi_{\rho}$ is then promoted to a self-adjoint Wilson loop operator $\hat{\chi}_{\rho}$ on $\mathcal{H}_{\mbox{\tiny{kin}}}$ creating loops in the $\rho$ representation around each plaquette defined by our regularization, which is charged for the face bounded by the loop $\alpha$. To summarize, we have, for each face $f$ of the regularization, a sum over the unitary, irreducible representations $\rho_{f}$ of $G$, a weight given by the dimension $\dim (\rho_f)$ of the representation $\rho_f$ summed over and a loop around the oriented boundary of the face in the representation $\rho_f$. See for instance \cite{karimale2}, \cite{thomasqsd},\cite{ashtekar} for details.
This concludes our regularization of the transition amplitudes of string-like sources coupled to four dimensional BF theory.

Now, the physical inner product that we have constructed above depends manifestly on the regulating structure $\Reg$. To complete the procedure, we have to calculate the limit in which the regulating parameters $\eta$ and $\epsilon$ go to zero.

\section{Regulator independence}

Throughout this section we will suppose that the cellular complex $(\T ,\partial \T)$ is simplicial, i.e. a triangulation of $(\M,T_{\eta})$. We will also make the simplifying assumption that $G=\SU(2)$, the unitary irreducible representations of which will be noted $(\stackrel{j}{\pi},\V_j)$, with the spin $j$ in $\N/2$. The generalization to arbitrary cellular decompositions and arbitrary compact Lie groups can be achieved by using the same techniques that we develop below.
If $\T$ is a now a regular triangulation, it cannot be adpated to any graph $\Gamma$. It can only be so for graphs with three-or-four-valent vertices. Now, we can always decompose a $n$-valent intertwiner, with $n > 3$, into a three-valent intertwiner by using repeatedly the complete reducibility of the tensor product of two representations
\footnote{One can show that the amplitudes obtained using the three-valent decomposition (where the virtual edges are assumed to be real) are identical to the ones obtained using a suitable non-simplicial cellular decomposition adapted to spin networks with arbitrary valence vertices.}.
We will therefore decompose all $n$-valent interwiners $\iota_v$ with $n > 3$ into three-valent vertices.

We now show how to remove the regulator $\Reg$ from the regularized scalar product \eqref{regulfinal1}. Instead of computing
the $\eta, \epsilon \rightarrow 0$ limit, we demonstrate that the transition amplitudes are in fact independent of the regulator. To prove such a statement, we show that the expression \eqref{regulfinal1} does not depend on any component of the regulator. The transition amplitudes are proven to be invariant under any finite combination of elementary moves called regulator moves $\EuScript{R} : \Reg \mapsto \mathcal{R}'_{(\eta,\epsilon')}$, where each regulator move is a combination of elementary moves acting on the components of the regulator :

\begin{itemize}
\item Bulk and boundary (adapted) Pachner moves $(\T, \partial \T) \mapsto (\T', \partial \T')$,
\item Elementary maximal tree moves $T \mapsto T'$,
\item Elementary curve moves $\gamma \mapsto \gamma'$.
\end{itemize}

We will see how the invariance under the above moves also implies
an invariance under dilatation/contraction of the string
thickening radius $\eta$.

To conclude on the topological invariance of the amplitudes from the above elementary moves, we will furthermore prove that the transition amplitudes are invariant under elementary moves acting on the string spin network graph $\EuScript{G}: \Gamma \mapsto \Gamma'$ which map ambient isotopic PL-graphs into ambient isotopic PL-graphs.

We now detail the regulator and graph topological moves.

\subsection{Elementary regulator moves}

The regulator moves are finite combinations of the following elementary moves acting on the simplical complex $(\T, \partial \T)$, the maximal tree $T$ and the paths $\alpha$ and $\beta \in C$.

\subsubsection{Adapted Pachner moves}

The first invariance property that we will need is that under moves acting on the simplicial-pair $(\T, \partial \T)$, leaving the one-complex $\Gamma$ {\itshape invariant} and mapping a $\Gamma$-adapted triangulation into a PL-homeomorphic $\Gamma$-adapted simplicial structure. We call these moves {\itshape adapted} Pachner moves. There are two types of moves to be considered : the bistellar moves \cite{bistellar}, acting on the bulk triangulation $\T$ and leaving the boundary simpicial structure unchanged, and the elementary shellings \cite{shellings}, deforming the boundary triangulation $\partial \T$ with induced action in the bulk.

\paragraph{Bistellar moves.}
There are four bistellar moves in three dimensions: the $(1,4)$, the $(2,3)$ and their inverses. In the first, one creates four tetrahedra out of one by placing a point $p$ in the interior of the original tetrahedron whose vertices are labeled $p_i$, $i=1,...,4$, and by adding the four edges $(p,p_i)$, the six triangles $(p,p_i,p_j)_{i \neq j}$, and the four tetrahedra $(p,p_i,p_j,p_k)_{i\neq j\neq k}$.
The $(2,3)$ move consists in the splitting of two tetrahedra into three : one replaces two tetrahedra $(u,p_1,p_2,p_3)$ and $(d,p_1,p_2,p_3)$ ($u$ and $d$ respectively refer to `up' and `down') glued along the $(p_1,p_2,p_3)$ triangle with the three tetrahedra $(u,d,p_i,p_j)_{i\neq j}$. The dual moves, that is the associated moves in the dual triangulation, follow immediately. See FIG. $2$.
\begin{figure}[h?]
\centerline{\hspace{0.5cm} \(
\begin{array}{ccc}
(4,1) & \hspace{5mm} & (3,2) \\
\includegraphics[width=4cm,height=1.6cm]{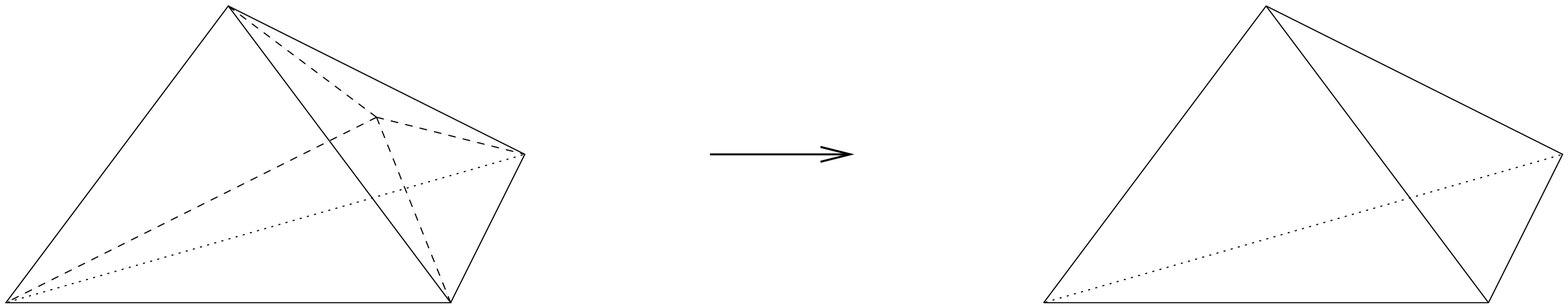} & \hspace{5mm} &
\includegraphics[width=3cm,height=1.6cm]{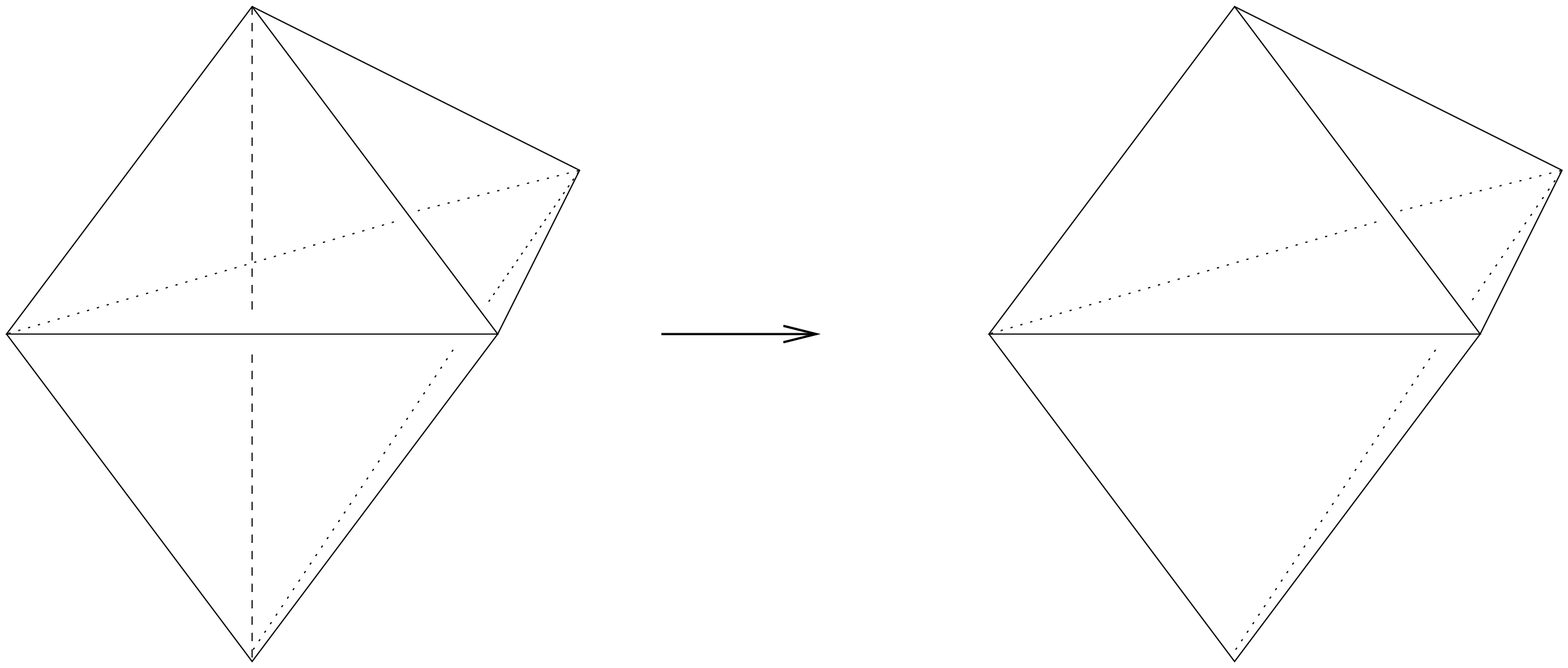} \\
\includegraphics[width=4cm,height=1.4cm]{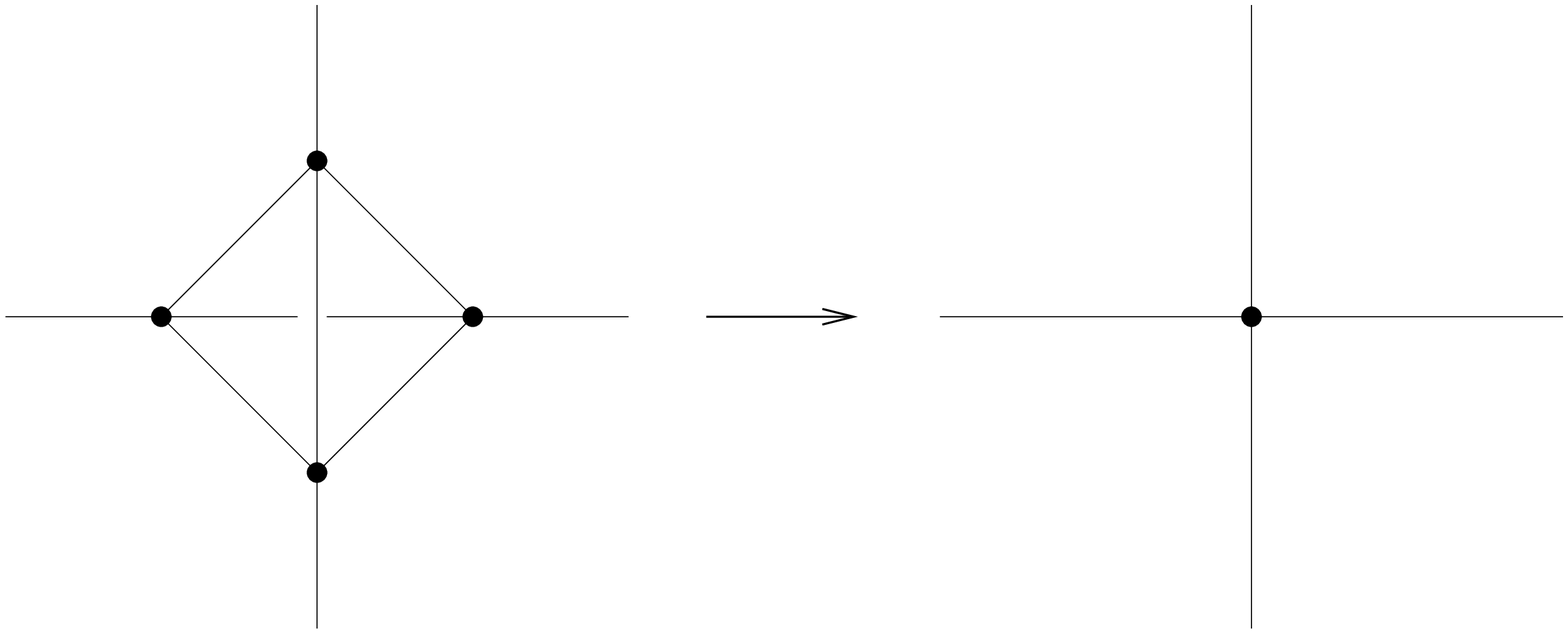} & \hspace{5mm} &
\includegraphics[width=3cm,height=1.4cm]{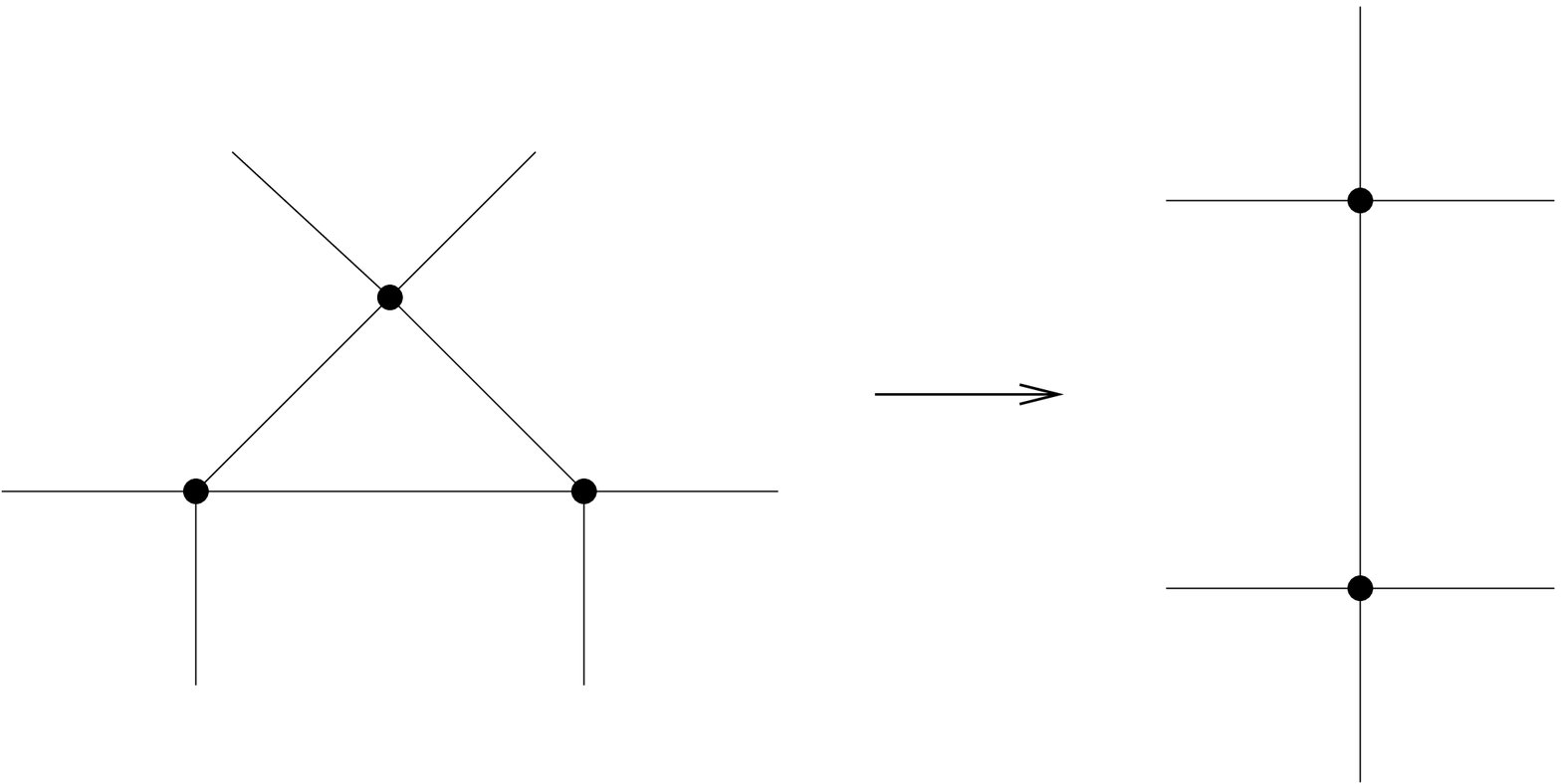}
\end{array}\)} \caption{The $(4,1)$ and $(3,2)$ bistellar moves.}
\end{figure}
\paragraph{Elementary shellings.}
Since the manifold $(\M, T_{\eta})$ has non-empty boundary, extra topological transformations have to be taken into account to prove discretization independence. These operations, called elementary shellings, involve
the cancellation of one $3$-simplex at a time in a given triangulation $(\Delta, \partial \Delta)$. In order to be deleted, the tetrahedron must have some of its two-dimensional faces lying in the boundary $\partial \Delta$. The idea is to remove three-simplices admitting boundary components such that the boundary triangulation admits, as new triangles after the move, the faces along which the given tetrahedron was glued to the bulk simplices. Moreover, for each elementary shelling there exists an inverse move which corresponds to the attachment of a new three-simplex to a suitable component in $\partial \T$. These moves correspond to bistellar moves on the boundary $\partial \Delta$ and there are accordingly three distinct moves for a three-manifold with boundary, the $(3,1)$, its inverse and the $(2,2)$ shellings, where the numbers $(p,q)$ here correspond to the number of two-simplices of a given tetrahedron lying on the boundary triangulation.
In the first, one considers a tetrahedron admitting three faces lying in $\partial \T$ and erases it such that the remaining boundary component is the unique triangle which did not belong to the boundary before the move. The inverse move follows immediately. The $(2,2)$ shelling consists in removing a three-simplex intersecting the boundary along two of its triangles such that, after the move, $\partial \T$ contains the two remaining faces of the given tetrahedron. These shellings and the associated boundary bistellars are depicted in FIG.$3$.
\begin{figure}[h?]
\centerline{\hspace{0.5cm} \(
\begin{array}{ccc}
(3,1) & \hspace{5mm} & (2,2) \\
\includegraphics[width=4cm,height=1.6cm]{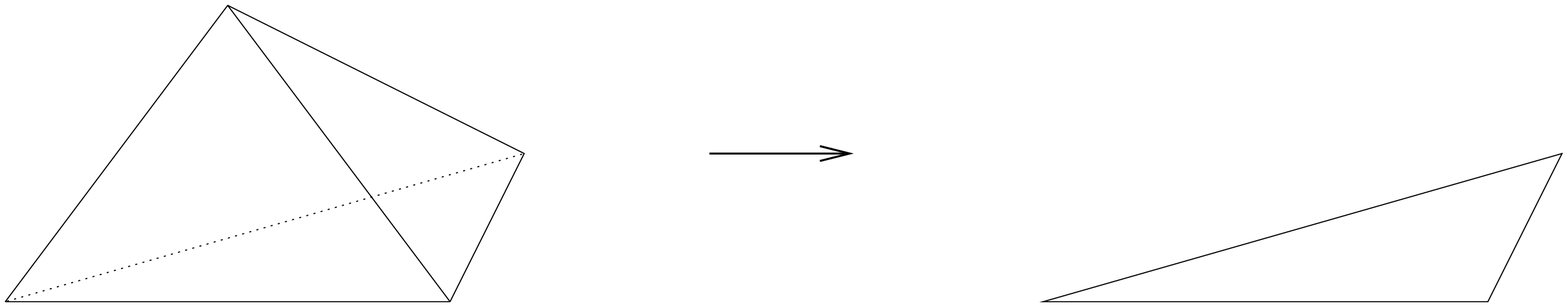} & \hspace{5mm} &
\includegraphics[width=3cm,height=1.6cm]{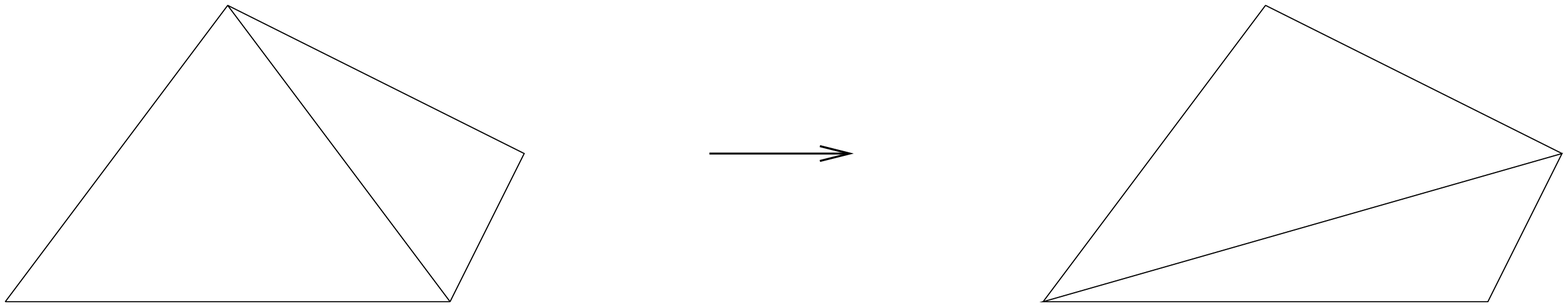}
\\
\includegraphics[width=4cm,height=1.4cm]{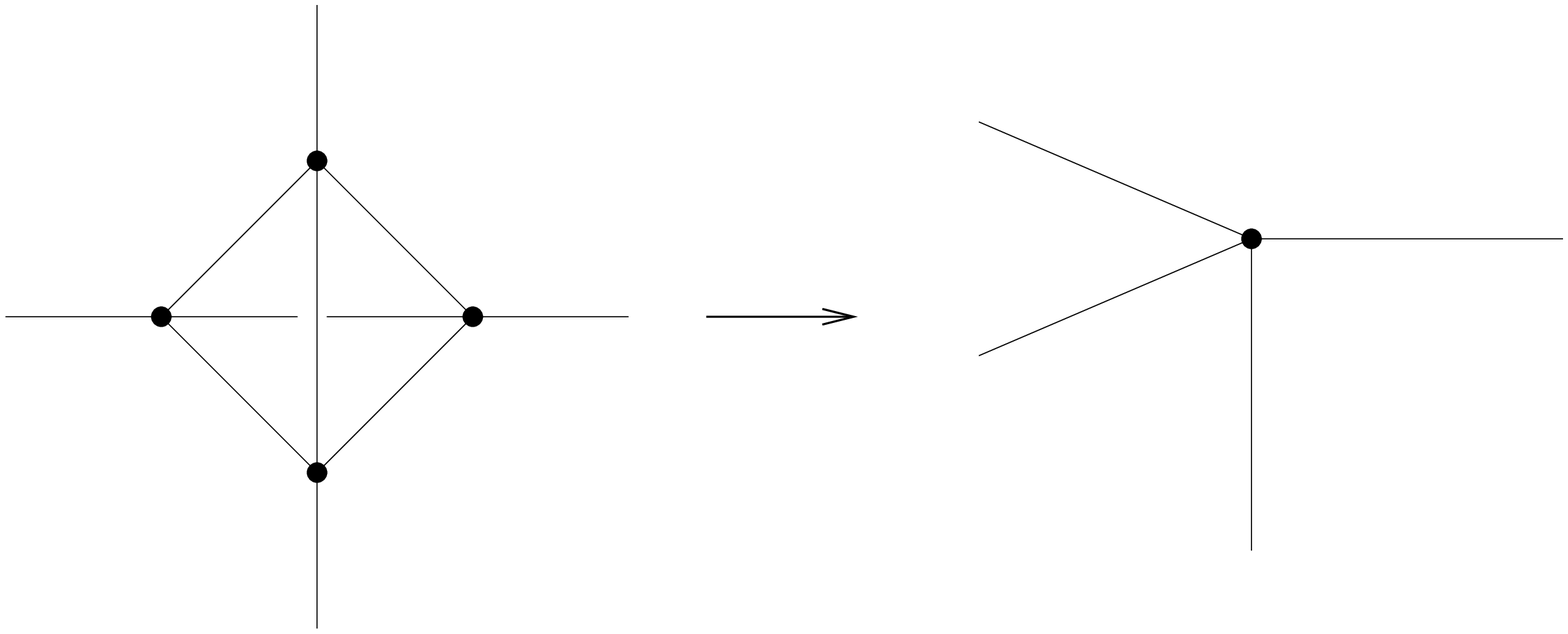} & \hspace{5mm} & \includegraphics[width=3cm,height=1.4cm]{dpachner3-2.eps}
\\
\includegraphics[width=4cm,height=1cm]{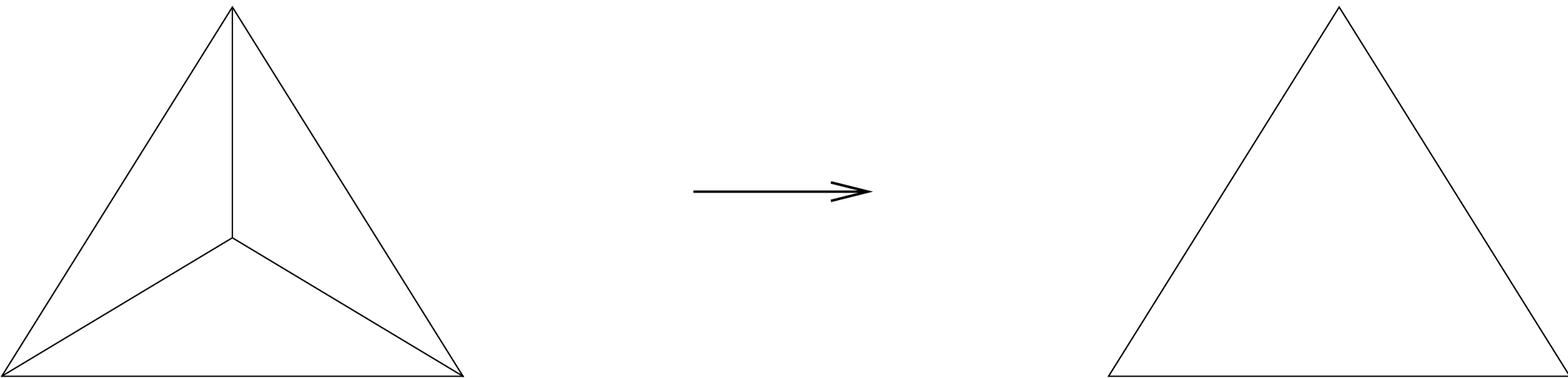} & \hspace{5mm} &
\includegraphics[width=3cm,height=1cm]{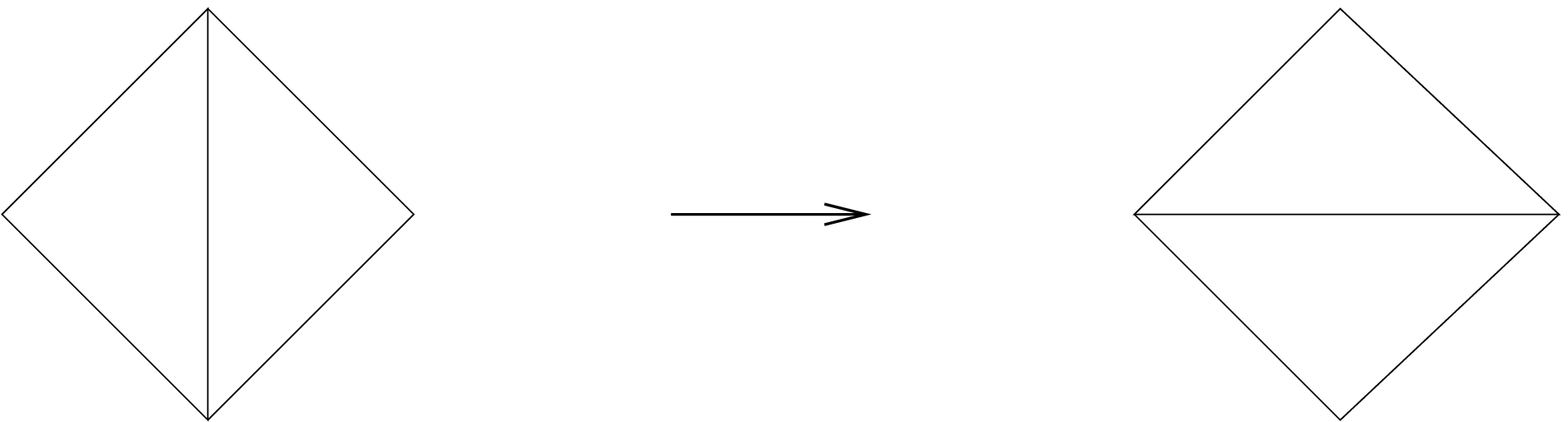}
\end{array}\)} \caption{The $(3,1)$ and $(2,2)$ shellings, their dual moves and the associated boundary bistellars.}
\end{figure}

The subset of bistellar moves and shellings which map $\Gamma$-adapted triangulations into $\Gamma$-adapted triangulations will be called adapted Pachner moves, and, considering a local simplicial structure $\mathcal{T}_k=\cup_{n=1}^k \T^n_3$, $k=1,...,4$, an adapted $(p,q)$ Pachner move $\mathcal{T}_p \mapsto \mathcal{T}_q$ will be noted $\EuScript{P}_{(p,q)}$, or more generally $\EuScript{P}$. Any two PL-homeomorphic, $\Gamma$-adapted triangulations $(\T,\partial \T)$ of the PL-pair $(\M,T_{\eta})$ are related by a finite sequence of such moves.

Here it is important to take into account the PL-embedding of the string spin network graph $\Gamma$ in the (dual) triangulation $(\T,\partial \T)$. We will call $\Gamma_{k} = \Gamma \cap \mathcal{T}_k^*$ the restrictions of the graph $\Gamma$ to the local simplex configurations $\mathcal{T}_k$ appearing in the moves.
If the graph $\Gamma_k$ is not the null graph, we will consider that it is open and does not contain any loop. If this was not the case, the set of  adapted moves would reduce to the identity move, under which the transition amplitudes are obviously invariant. Hence, $\Gamma_k$, if at all, can only be either an edge, or more generally a collection of edges, either a (three-valent) vertex.
The associated string spin network functional $\Psi_{\Gamma_k}$ will be represented by a group function $\phi_k$, which is the constant map $\phi_k=1$ if $\Gamma_k$ is the null graph.
We will make sure to check that, under a $(p,q)$ Pachner move, $\phi_k$ transforms as $\phi_p \mapsto \phi_q$.

\subsubsection{Maximal tree moves}

It is also necessary to define topological moves for the trees \cite{etera}, \cite{pr2}. Any two homologous trees $T_1$ and $T_2$ are related by a finite sequence of the following elementary tree moves $\EuScript{T} : T_1 \mapsto T_2$.

\begin{df}
{\bf (Tree move)} Considering a vertex $\T_0$ belonging to a tree $T$, choose a pair of edges $\Delta_1, \Delta_1'$ in $\T$ touching the vertex $\Delta_0$ such that $\Delta_1$ is in $T$, $\Delta_1'$ is not in $T$ and such that $\Delta_1'$ combined to the other edges of $T$ does not form a loop. The move $\EuScript{T}$ consists in erasing the edge $\Delta_1$ from $T$ and replacing it by $\Delta_1'$.
\end{df}

\begin{figure}[h?]
\centerline{\hspace{0.5cm} \(
\psfrag{a}{$\Delta_0$}
\psfrag{b}{$\Delta_1$}
\psfrag{c}{$\Delta_1'$}
\includegraphics[width=10cm,height=4cm]{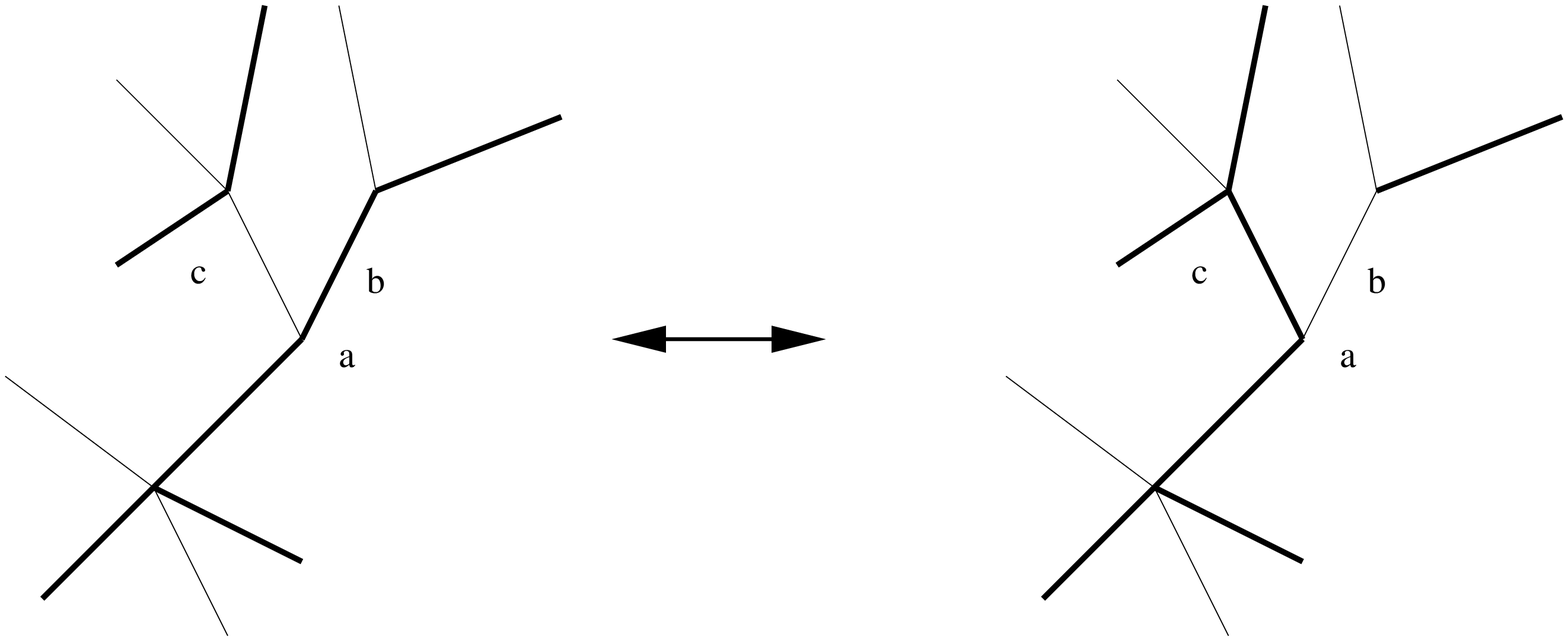} \)}
\end{figure}

There is another operation on trees that we need to define. When acting on the simplicial complex $(\Delta,\partial \T)$ with a bistellar move or a shelling, one can possibly map $(\Delta, \partial \T)$ into a simplicial complex $(\Delta',\partial \T')$ with a different number of vertices. Hence, a maximal tree $T$ of $\Delta$ is not necessarily a maximal tree of $\Delta'$; the Pachner moves have a residual action on the trees. This leads us to define the notion of maximal tree extension (or reduction) accompanying Pachner moves modifying the number of vertices of the associated simplicial complex.

\begin{df}
{\bf (Tree extension or reduction)}
An extended (or reduced) tree $T$, associated to a Pachner move $\EuScript{P} : \Delta \mapsto \Delta'$ modifying locally the number of vertices of a simplicial complex $\Delta$, is a maximal tree of $\Delta'$ obtained from a maximal tree $T$ of $\Delta$ by adding (or removing) the appropriate number of edges to $T$ as a mean to transform $T$ into a maximal tree $T_{\EuScript{P}}$ of $\Delta'$.
\end{df}

Obviously, there is an ambiguity in the operation of tree extension or reduction. But, because of the fact that the regularized physical inner product will turn out to be independent of a choice of maximal tree, there will be no trace of this ambiguity in the computations of the transition amplitudes.

\subsubsection{Curve moves}

Finally, we define the PL analogue of the Reidemeister moves, which where in fact a crucial ingredient in the proof of Reidemeister's theorem. Any two ambient isotopic PL embeddings $\gamma_1$ and $\gamma_2$ of a curve $\gamma$ in the dual complex $(\T^*,\partial \T^*)$ are related by a finite sequence of the following elementary topological moves $\EuScript{C} : \gamma_1 \mapsto \gamma_2$.

\begin{df}
{\bf (Curve move)} Consider a PL path $\gamma$ lying along the $p$ boundary edges $e_1, e_2, ..., e_p$ of a two-cell $f$ of the dual pair $(\T^*,\partial \T^*)$, where $f$ has no other edges nor vertices traversed by the curve $\gamma$. Erase the path $\gamma$ along the edges $e_1, e_2, ..., e_p$ and add a new curve along the complementary $\partial f \setminus \{ e_1, e_2, ..., e_p \}$ of the erased segment in $f$.
\end{df}

\begin{figure}[h?]
\centerline{\hspace{0.5cm} \(
\psfrag{a}{$e$}
\psfrag{b}{$\partial f \setminus e$}
\psfrag{d}{$f$}
\includegraphics[width=7cm,height=2.5cm]{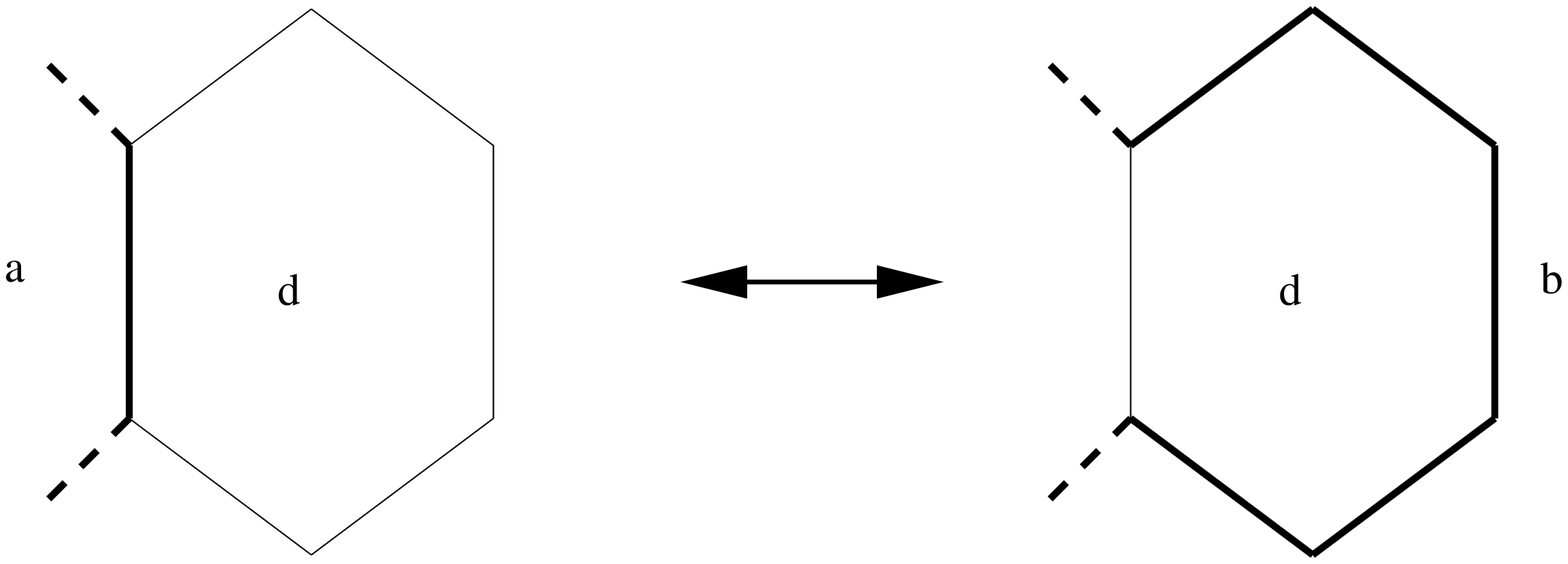} \)}
\end{figure}

We have now defined all of the elementary regulator moves. To
summarize, an elementary regulator move $\EuScript{R} : \Reg
\mapsto \mathcal{R}'_{(\eta,\epsilon')}$ is a finite combination
of all the above moves : $\EuScript{R}[\Reg] = (T_{\eta},
\EuScript{P}[(\T,\partial \T)], \EuScript{T}[T_{\EuScript{P}}],
\EuScript{C}[\alpha], \EuScript{C}[C])$. Proving the invariance of
the regularized physical inner product \eqref{regulfinal1} under
all of these elementary regulator moves is equivalent to showing
the independence of the regulating structure $\Reg$. Note however
that we have not included contractions or dilatations of the
string tube $T_{\eta}$ radius $\eta$ in the regulator moves. This
is because the invariance under shellings implies the invariance
under increasing or decreasing of $\eta$. Indeed, the bistellars
and shellings are the simplicial analogues of the action of the
homeomorphisms Homeo$[(\M,T_{\eta})]$. In particular, the
topological group Homeo$[(\M,T_{\eta})]$ contains transformations
deforming continuously the boundary $T_{\eta}$, like for instance
maps decreasing or increasing the (non-contractible) radius
$\eta>0$ of the boundary torus $T_{\eta}$. Hence, showing the
invariance under elementary shellings is sufficient to prove the
independence on the string thinckening radius, and the moves
defined above are sufficient to conclude on the regulator
independence of the definition of the regularized physical inner
product.

To push the result further and conclude on the topological invariance of the transition amplitudes, we need extra ingredients that we define in the following section.

\subsection{String spin network graph moves}

We now introduce the following elementary moves respectively acting on the edges and vertices of the open graph $\Gamma$.
All ambient isotopic PL embeddings of the one complex $\Gamma$ are related by a finite sequence of the following elementary moves noted $\EuScript{G}$.

\begin{df}
{\bf (Edge move)} An edge move is a curve move applied to an edge $e_{\Gamma}$ of the graph $\Gamma$.
\end{df}

Note that these moves apply also to the open edges of the graph $\Gamma$. However, there exists other moves which displace the endpoints.

\begin{df}
{\bf (Endpoint move)} Considering an open string spin network edge $e_{\Gamma}$ ending on the point $x_k \in X$ supported by a dual vertex $\overline{v}$ of $\partial \T^*$, which is such that its neighbouring vertex $\overline{v}'$ not touched by $e_{\Gamma}$ belongs to $\partial \T^*$, an endpoint move consist in adding a section to $e_{\Gamma}$ connecting $\overline{v}$ to $\overline{v}'$.
\end{df}

\begin{figure}[h?]
\centerline{\hspace{0.5cm} \(
\psfrag{a}{$e$}
\psfrag{b}{$\partial f \setminus e$}
\psfrag{c}{$\overline{v}$}
\psfrag{d}{$f$}
\psfrag{e}{$\overline{v}'$}
\includegraphics[width=7cm,height=2.5cm]{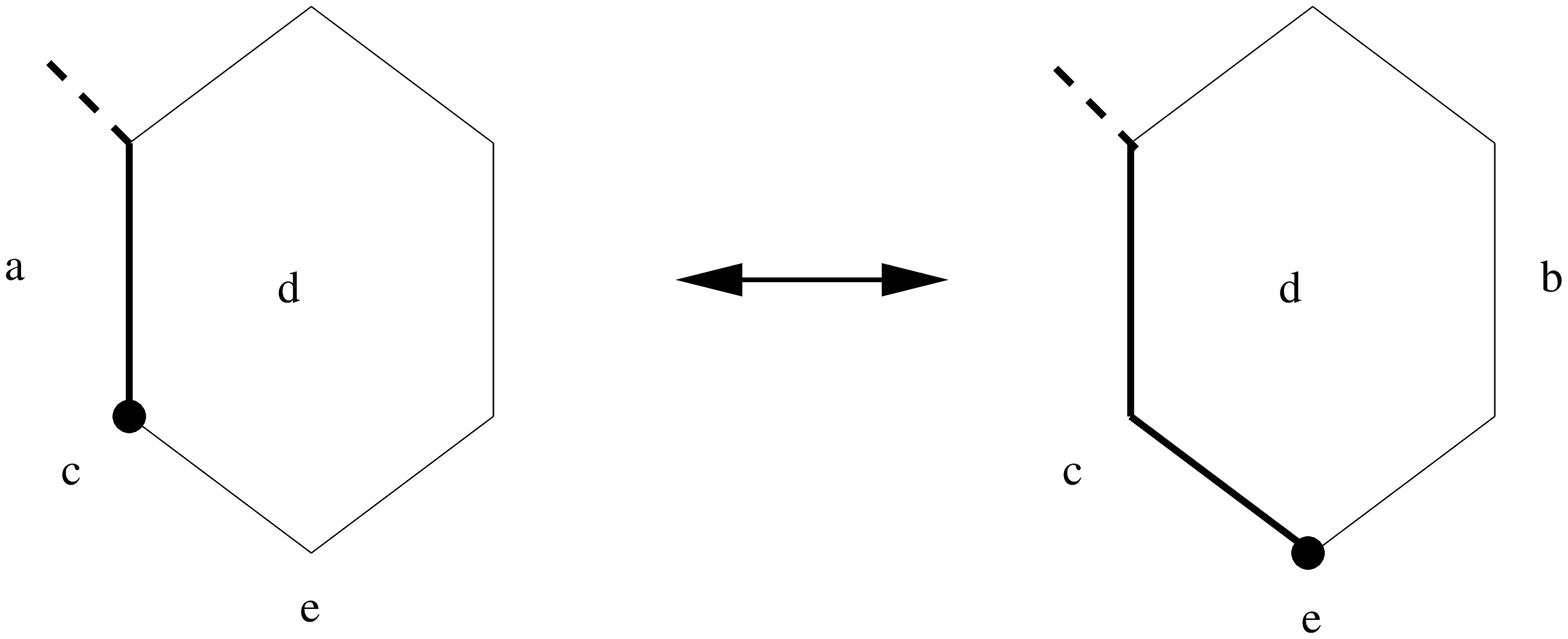} \)}
\end{figure}

We also need similar moves for the vertices.

\begin{df}
{\bf (Vertex translation)} Let $v_{\Gamma}$ denote a three-valent spin network vertex sitting on the vertex $v$ of the dual complex $(\T^*, \partial \T^*)$. Choose one edge $e_{\Gamma}$ among the three edges emerging from $v_{\Gamma}$ and call $v'$ the dual vertex adjacent to $v$ which is traversed by $e_{\Gamma}$. Call $e,e' \subset (\T^*, \partial \T^*)$ the dual edges locally supporting $e_{\Gamma}$, i.e, such that $\partial e = \{ v, v' \}$ and $v' = e_{\Gamma} \cap (e \cap e')$.

The move consists in translating the vertex $v_{\Gamma}$ along $e$ from $v$ to $v'$. This is achieved by choosing one dual face sharing the dual edge $e$ and not containing the dual edge $e'$, and acting upon it with the edge move.
\end{df}

\begin{figure}[h?]
\centerline{\hspace{0.5cm} \(
\psfrag{a}{$v$}
\psfrag{b}{$v'$}
\psfrag{c}{$e$}
\psfrag{d}{$e'$}
\includegraphics[width=12cm,height=3.6cm]{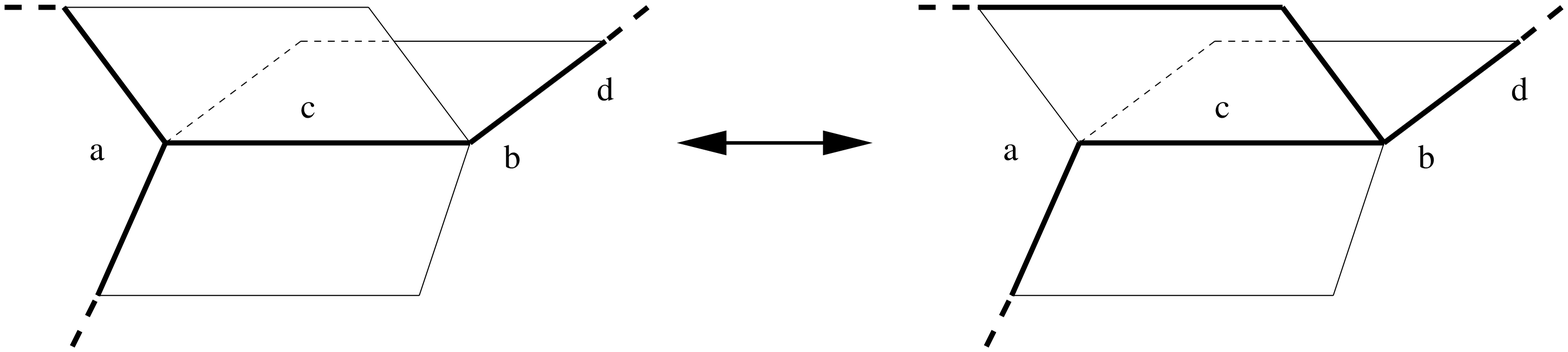} \)}
\end{figure}

Note that the use of rectangular faces in the above picture is only for the clarity of the picture, the move is defined for faces of arbitrary shape.

It is important to remark that the above moves respect the topological structure of the embedding because no discontinuous transformations are allowed and because the number and nature of the crossings are preserved since the faces used to define the moves are required to have empty intersections with the string or graph a part from the specified ones.

The combination of the adapted Pachner moves and the spin network moves are the simplicial analogues of the action of the homeomorphisms Homeo$[(\M,T_{\eta})]$ on the triple $((\M,T_{\eta}),\Gamma)$.

\subsection{Invariance theorem}

We can now prove the following theorem.

\begin{Th}
{\bf (Invariance theorem)}
Let $\Psi_{\Gamma}$ denote a string spin network element of a given basis of $\mathcal{H}_{kin}$ defined with respect to the one-complex $\Gamma$. Choose a regulator $\mathcal{R}_{(\eta,\epsilon)} = (T_{\eta}, (\T_{\epsilon},\partial \T_{\epsilon}), T, \alpha, C)$ consisting in a thickening $T_{\eta}$ of the string, a cellular decomposition $(\T_{\epsilon},\partial \T_{\epsilon})$ of the manifold $(\M,T_{\eta})$ adapted to the graph $\Gamma$, a maximal tree $T$ of $\T$, a closed path $\alpha$ in $\partial \T^*$, and a collection $C$ of open paths $\beta$ in $\partial \T^*$. Let $\EuScript{R} : \Reg \mapsto \mathcal{R}'_{(\eta,\epsilon')}$ (resp. $\EuScript{G}: \Gamma \mapsto \Gamma'$) denote an elementary regulator move (resp. a string spin network move). The evaluated linear form \eqref{regulfinal2} is invariant under the action of $\EuScript{R}$ and $\EuScript{G}$:
\beqa
P[\mathcal{R}_{(\eta,\epsilon)};\Psi_{\Gamma}] &=& P[\, \EuScript{R}[\Reg] ; \Psi_{\Gamma}] \label{th1} \\ \nn
                                               &=& P[\Reg ; \Psi_{\EuScript{G}(\Gamma)}]. \label{th2}
\eeqa
\end{Th}

{\itshape Proof.} We proceed by separatly showing the invariance under each elementary regulator moves, before proving the invariance under graph moves.

\begin{itemize}
\item {\em Invariance under maximal tree moves.}

Here, we simply apply to the proof of invariance under maximal tree moves writen in \cite{pr2}.
Firstly, we need to endow the tree $T$ of the left hand side of the move with a partial order. To this aim, we pick a distinguished vertex $r$ of $T$, chosen to be the other vertex of the edge $\T_1'$. The rooted tree $(T,r)$ thus acquires a partial order $\preceq$ : a vertex $\T_0'$ of $(T,r)$ is {\itshape under} a vertex $\T_0$, $\T_0' \preceq \T_0$, if it lies on the unique path connecting $r$ to $\T_0$. We can now define the tree $T_{\T_0}$ to be the subgraph of $T$ connecting all the vertices {\itshape above} $\T_0$ : $\T_0$ is the root of $T_{\T_0}$.
The second ingredient that we need is the notion of Bianchi identity \eqref{bianchi} applied to trees. Indeed, if $T$ is a tree of the (regular) simplicial complex $\T$, its tubular neighborhood has the topology of a $3$-ball and its boundary has the topology of a $2$-sphere. This surface $\mathcal{S}$ can be built as the union of the faces $f$ dual to the edges $\Delta_1$ in $\T$ touching the vertices of $T$ without belonging to $T$.
Hence, applying the Bianchi identity to the tree $T_{\T_0}$ yields
\beq
g_{f_0'}= \left( \prod_{f \in \mathcal{S}_1} g_f \right) \, g_{f_0} \, \left( \prod_{f \in \mathcal{S}_2} g_f \right).
\eeq
Here, $f_0$ and $f'_0$ are the faces dual to the segments $\T_0$ and $\T_0'$ (note that $\T_0$ does not belong to $T_{\T_0}$). The sets $\mathcal{S}_1$, $\mathcal{S}_2$ are the set of faces dual to the segments $\T_1$ touching the vertices of $T_{\T_0}$ without belonging to $T_{\T_0}$, and which are not $f_0$ nor $f_0'$. The presence of two different sets $\mathcal{S}_1$ and $\mathcal{S}_2$ is simply to take into account the arbitrary positioning of the group element $g_{f_0}$ amoung the product over all faces. As usual, the group elements are defined up to orientation and conjugation.
Next, we apply a delta function to both sides of the above equation and multiply the result as follows :
\beqa
\prod_{{\tiny \begin{array}{cc} f \notin T \\ f \neq f_0,f_0' \end{array}}} \delta(g_f) \,\, \delta(g_{f_0'})
&=& \prod_{{\tiny \begin{array}{cc} f \notin T \\ f \neq f_0,f_0' \end{array}}} \delta(g_f) \,\, \delta \left[ \left( \prod_{f \in \mathcal{S}_1} g_f \right) \, g_{f_0} \, \left( \prod_{f \in \mathcal{S}_2} g_f \right) \right] \\ \nn
\prod_{f \notin T} \delta(g_f) &=& \prod_{f \notin T'} \delta(g_f).
\eeqa
In the second step, we have simply used the delta functions with which the expression has been multiplied to
set the group elements associated to the sets $\mathcal{S}_1$ and $\mathcal{S}_2$ to the identity (the faces of $\mathcal{S}_1$ and $\mathcal{S}_2$ are dual to segments not belonging to $T$). One can then check that the various steps of the proof remain valid if the boundaries of the dual faces carry string spin networks. This shows the invariance of the regularized inner product \eqref{regulfinal2} under maximal tree move ${\tiny \square}$

\item {\em Invariance under adapted Pachner moves.}

To prove the invariance under Pachner moves, we introduce a simplifying lemma \cite{robert}, \cite{arf}.

\begin{Lm}
{\bf (Gauge fixing identity)}
To each vertex of the dual triangulation $\T^*$ are associated four group elements $\{g_a\}_{a=1,...,4}$, six unitary irreducible representations $\{j_{ab}\}_{a<b=1,...4}$ of $G$ and a string spin network function $\phi_1(\{g_a\}_{a=1,...,4})$.
If $\phi_1$ is the constant map $\phi_1=1$, or depends on its group arguments only through monomial combinations $\{g_ag_b\}_{a\neq b}$ of degree two, then the following identity holds :
\beq
\prod_{b>a=1}^4 \int_G dg_a \stackrel{j{ab}}{\pi}(g_a g_b) \, \, \phi_1(\{g_d\}_d) = \prod_{b>a=1}^4 \int_G dg_a \, \delta(g_c) \, \stackrel{j{ab}}{\pi}(g_a g_b) \, \, \phi_1(\{g_d\}_d),
\eeq
for $c=1,2,3$ or $4$.
\end{Lm}

{\itshape Proof of Lemma $1$.} The above equality is trivially proven by using the invariance of the Haar measure and performing the change of variables $\gamma_{cb} = g_c g_b$, for $c<b$ (resp. $\gamma_{bc} = g_b g_c$, for $c>b$) in the left hand side. This translation is always possible since the group function $\phi$ is either the constant map or depends on the group elements only through monomials of degree two {\small $\square$}

Let us comment here on the validity of the hypothesis made on the spin network function $\phi_1$ associated to the graph $\Gamma_1 = \Gamma \cap \mathcal{T}_1$, with $\mathcal{T}_1 = \T_3$ in the above Lemma.
In fact $\phi_1$ depends necessarily on combinations of the form $g_ag_b$ locally if the graph $\Gamma_1$ is not the null graph. Indeed, $\Gamma_{1}$ can either be a collection of edges, in which case this requirement simply states that the edges are open, either a vertex, where one can always use the invariance of the associated intertwining operator to satisfy the desired assumption. Hence, this requirement is always locally satisfied.

We can now show the invariance under bistellar moves and shellings.

\begin{itemize}
\item Bistellars moves:
\begin{itemize}
\item {\bf The $(4,1)$ move.} Consider the four simplices configuration $\mathcal{T}_4$ in $(\T,\partial \T)$ (FIG. $2$). Since the amplitudes do not depend on the maximal tree $T$ of $\T$, we are free to chose it. The simplest choice consists in considering a maximal tree $T$ whose intersection $T_4$ with the simplex configuration $\mathcal{T}_4$ reduces to the four external vertices and to a single one-simplex touching the central vertex. We work in the dual picture and label the four external dual edges from one to four. The face dual to the internal tree segment is chosen to be the face $142$. We note $g_a$ and $h_{ab}$, $a,b=1,...,4$, the group elements \footnote{The notations takes into account the appropriate orientations.} associated to the external and internal dual edges respectively, while the representations assigned to the dual faces are noted $j_{ab}$. The general PL string spin network state restricted to the configuration $\mathcal{T}_4^*$ is noted $\phi_4(\{g_a\}_a,\{h_{ab}\}_{a<b})$. Obviously, $\phi_4$ is not a function of all of its ten arguments, otherwise it would contain a loop, but can generally depend on any one of these ten group elements, as suggested by the notation. The regularized physical inner product \eqref{regulfinal2} restricted to these four simplices yields
\beqa
&& \int_{G^{10}} dg_1 dg_2 dg_3 dg_4  \, d h_{12} d h_{13} d h_{14} d h_{23} d h_{24} d h_{34} \\ \nn
&& \stackrel{j_{12}}{\pi} (g_1 h_{12} g_2) \stackrel{j_{13}}{\pi} (g_1 h_{13} g_3) \stackrel{j_{14}}{\pi} (g_1 h_{14} g_4) \stackrel{j_{23}}{\pi} (g_2 h_{23} g_3) \stackrel{j_{24}}{\pi} (g_2 h_{24} g_4) \stackrel{j_{34}}{\pi} (g_3 h_{34} g_4) \\ \nn
&& \delta(h_{12} h_{23} h_{31}) \, \delta(h_{13} h_{34} h_{41}) \, \delta(h_{23} h_{34} h_{24}) \,\, \phi_4(\{g_a\}_a,\{h_{ab}\}_{a<b}),
\eeqa
where we have omitted the sum over representations weighted by the associated dimensions.
We start by implementing Lemma $1$ at vertices $2$, $3$ and $4$ to eliminate the three variables $h_{1b}$, $b \neq 1$. We obtain
\beqa
&& \int_{G^{7}} dg_1 dg_2 dg_3 dg_4  \, d h_{23} d h_{24} d h_{34} \\ \nn
&& \stackrel{j_{12}}{\pi} (g_1 g_2) \stackrel{j_{13}}{\pi} (g_1 g_3) \stackrel{j_{14}}{\pi} (g_1 g_4) \stackrel{j_{23}}{\pi} (g_2 h_{23} g_3) \stackrel{j_{24}}{\pi} (g_2 h_{24} g_4) \stackrel{j_{34}}{\pi} (g_3 h_{34} g_4) \\ \nn
&& \delta(h_{23}) \, \delta(h_{34}) \, \delta(h_{23} h_{34} h_{24}) \,\, \phi(\{g_a\}_a,\{h_{23},h_{34},h_{24}\}) \\ \nn
&=& \int_{G^{4}} dg_1 dg_2 dg_3 dg_4 \stackrel{j_{12}}{\pi} (g_1 g_2) \stackrel{j_{13}}{\pi} (g_1 g_3) \stackrel{j_{14}}{\pi} (g_1 g_4)
\\ \nn
&& \stackrel{j_{23}}{\pi} (g_2 g_3) \stackrel{j_{24}}{\pi} (g_2 g_4) \stackrel{j_{34}}{\pi} (g_3 g_4) \, \, \phi_1(\{g_a\}_a),
\eeqa
where we have integrated over the delta functions to eliminate the interior variables in the second step. The right hand side of the above equality corresponds to the one simplex configuration of the $(4,1)$ move with the associated maximal tree reduction (the obvious removal of the internal tree segment; $T_1$ is given by the four vertices of the resulting tetrahedron). In other words, we have just proved the invariance under the transformation $\EuScript{P}_{(4,1)} : (\mathcal{T}_4,T_4) \mapsto (\mathcal{T}_1,T'_1)$, with $T_k = T \cap \mathcal{T}_k$ and $T'=T_{\EuScript{P}_{(4,1)}}$.

\item {\bf The $(3,2)$ move.} Here, we consider the three simplices configuration $\mathcal{T}_3$ (see FIG. $2$) and chose a tree $T$ intersecting $\mathcal{T}_3$ only on its five vertices. Concentrating on the dual graph, we label the three vertices from one to three and respectively note $g_a^{\alpha}$, $h_{ab}$, and $j_{ab}^{\alpha \beta}$, $a,b=1,...,3$, $\alpha,\beta=1,2$, the external and internal group elements, and the representation labels. The associated string spin network state is called $\phi_3$. The transition amplitude, restricted to these three simplices, yields (omitting the sum over representations and associated dimensions)
\beqa
&& \int_{G^{9}} dg_1^1 dg_1^2 dg_2^1 dg_2^2 dg_3^1 dg_3^2  \, d h_{12} d h_{13} d h_{23} \\ \nn
&& \stackrel{j_{12}^{11}}{\pi} (g_1^1 h_{12} g_2^1) \stackrel{j_{13}^{11}}{\pi} (g_1^1 h_{13} g_3^1) \stackrel{j_{23}^{11}}{\pi} (g_2^1 h_{23} g_3^1) \stackrel{j_{12}^{22}}{\pi} (g_1^2 h_{12} g_2^2 ) \stackrel{j_{13}^{22}}{\pi} (g_2^1 h_{13} g_3^2) \stackrel{j_{23}^{22}}{\pi} (g_2^2 h_{23} g_3^2)\\ \nn
&& \stackrel{j_{11}^{12}}{\pi} (g_1^1 g_1^2) \, \stackrel{j_{22}^{12}}{\pi} (g_2^1 g_2^2)\, \stackrel{j_{33}^{12}}{\pi} (g_3^1 g_3^2)\, \delta(h_{12} h_{23} h_{13}) \,\, \phi_3(\{g_a\}_a,\{h_{ab}\}_{a<b}).
\eeqa
Using the gauge fixing identity at the vertices $2$ and $3$ to eliminate the variables $h_{1a}$, $a \neq 1$, and solving for the delta function leads to
\beqa
&& \int_{G^{6}} dg_1^1 dg_1^2 dg_2^1 dg_2^2 dg_3^1 dg_3^2 \\ \nn
&& \stackrel{j_{12}^{11}}{\pi} (g_1^1 g_2^1) \stackrel{j_{13}^{11}}{\pi} (g_1^1 g_3^1) \stackrel{j_{23}^{11}}{\pi} (g_2^1 g_3^1) \stackrel{j_{12}^{22}}{\pi} (g_1^2 g_2^2 ) \stackrel{j_{13}^{22}}{\pi} (g_2^2 g_3^2) \stackrel{j_{23}^{22}}{\pi} (g_2^2  g_3^2)\\ \nn
&& \stackrel{j_{11}^{12}}{\pi} (g_1^1 g_1^2) \, \stackrel{j_{22}^{12}}{\pi} (g_2^1 g_2^2)\, \stackrel{j_{33}^{12}}{\pi} (g_3^1 g_3^2) \, \, \phi(\{g_a\}_a) \\ \nn
&=& \int_{G^{7}} dg_1^1 dg_1^2 dg_2^1 dg_2^2 dg_3^1 dg_3^2  \, d h \\ \nn
&& \stackrel{j_{12}^{11}}{\pi} (g_1^1 g_2^1) \stackrel{j_{13}^{11}}{\pi} (g_1^1 g_3^1) \stackrel{j_{23}^{11}}{\pi} (g_2^1 g_3^1) \stackrel{j_{12}^{22}}{\pi} (g_1^2 g_2^2 ) \stackrel{j_{13}^{22}}{\pi} (g_2^2 g_3^2) \stackrel{j_{23}^{22}}{\pi} (g_2^2  g_3^2)\\ \nn
&& \stackrel{j_{11}^{12}}{\pi} (g_1^1 h g_1^2) \, \stackrel{j_{22}^{12}}{\pi} (g_2^1 h g_2^2)\, \stackrel{j_{33}^{12}}{\pi} (g_3^1 h g_3^2) \, \, \phi_2(\{g_a\}_a,h),
\eeqa
where we have used the inverse gauge fixing identity in the last step. This expression corresponds to the two simplices configuration $\mathcal{T}_2$ of the $(3,2)$ move.
\end{itemize}

\item Shellings :
\begin{itemize}
\item {\bf The $(3,1)$ move.} Remarkably, writting the amplitudes associated to the left and right hand sides of the $(3,1)$ shelling leads to the same expression than the $(4,1)$ bistellar, even if the geometrical interpretation is obviously different. This is due to the fact that we are imposing the flatness constraint $F=0$ also on the faces of the boundary  \footnote{In this sense, the boundary amplitudes are very different from a $2+1$ quantum gravity model defined on an open manifold.} simplicial complex $\partial \T$ and integrating also on the boundary edges. The only difference is in the presence of possible open string spin network edges reflected in the group function $\phi = \phi(\{g_a\}_a,\{\lambda_a\}_a)$, without any incidence an any steps of the proof given for the $(4,1)$ bistellar.
Accordingly, the proof of invariance under the $(3,1)$ shelling is the one sketched above.

\item {\bf The $(2,2)$ move.} The same remark applies here, the amplitudes are exactly identical to the ones of the $(3,2)$ bistellar.
\end{itemize}

Accordingly, we have proven the invariance under adapted Pachner moves ${\tiny \square}$
\end{itemize}

\item {\em Invariance under curve moves.}
We here show that the regularized physical inner product is invariant under curve moves. The proof uses the flatness constraint $F=0$.
Consider a particular dual face $f$ of $(\T^*, \partial \T^*)$ containing $n$ boundary edges positively oriented from vertex $1$ to vertex $n$. Suppose that there are $p < n$ dual edges $e_1,...,e_p$ supporting a curve positively oriented w.r.t the face $f$, to which is associated a spin $j$ representation. We want to prove that the associated amplitude is equal to the amplitude corresponding to the curve lying along the $n-p$ edges of $\partial f \setminus \{ e_1,...,e_p \}$ after the edge move. We start from the initial configuration
\beq
\int_{G^n} \prod_{a=1}^n dg_a \stackrel{j}{\pi} (g_1 ... g_p) \, \delta(g_1 ... g_p g_{p+1} ... g_n) \, \delta(g_1 G_1) \delta(g_1 H_1) \, ... \, \delta(g_n G_n) \, \delta(g_n H_n),
\eeq
where the capital letter $G_a,H_a$, $a=1,...,n$, represent the sequences of group elements associated to the two others faces sharing the edge $a$. We then simply integrate over the group element $g_1$ to obtain
\beqa
&& \int_{G^{n-1}} \prod_{a=2}^n dg_a \stackrel{j}{\pi} (g_n^{-1} ... g_{p+1}^{-1}) \, \delta(g_n^{-1}...g_2^{-1} G_1) \, \delta(g_n^{-1}...g_2^{-1} H_1) \, ... \, \delta(g_n G_n) \, \delta(g_n H_n) \\ \nn
&=& \int_{G^n} \prod_{a=1}^n dg_a \stackrel{j}{\pi} (g_n^{-1} ... g_{p+1}^{-1}) \, \delta(g_1 ... g_p g_{p+1} ... g_n) \, \delta(g_1 G_1) \, \delta(g_1 H_1) \, ... \, \delta(g_n G_n) \, \delta(g_n H_n).
\eeqa
Note the reversal of orientations intrinsic to the move. This closes the proof of invariance under curve move ${\tiny \square}$

We finish the proof of theorem $1$ by showing the second part, namely the invariance of the transition amplitudes under string spin network graph moves.

\item {\em Invariance under edge moves.}

The proof is the one given for the curve move ${\tiny \square}$

\item {\em Invariance under endpoint moves.}

Here, we use the momentum conservation $Dp=0$.
Considering a particular dual face $f$ containing $n$ boundary edges, with $p < n$ dual edges $e_1,...,e_p$ supporting an open string spin network edge (positively oriented w.r.t $f$) ending on the boundary of the edge $p$, to which is associated a spin $j$ representation. We call $\lambda_k$ the string field evaluated at the target of the $k$th edge. We choose the holonomy starting point $x$ to be on the endpoint of the $p$ edge (we prove below that nothing depends on this choice) and, since nothing depends on the paths $\beta$ by virtue of the invariance under curve moves, we choose a path $\beta$ of $C$ along the edge $p+1$. The relevant amplitude is given by
\beq
\int_{G^n} \prod_{a=1}^n dg_a d\lambda_{p} d \lambda_{p+1} \stackrel{j}{\pi} (g_1 ... g_p \lambda_p) \delta(g_{p+1} \lambda_{p+1} \lambda_p^{-1}),
\eeq
where the notations are the same than above. It is immediate to rewrite the above quantity as
\beq
\int_{G^n} \prod_{a=1}^n dg_a d\lambda_{p} d \lambda_{p+1} \stackrel{j}{\pi} (g_1 ... g_p g_{p+1} \lambda_{p+1}) \delta(g_{p+1} \lambda_{p+1} \lambda_p^{-1}),
\eeq
which concludes the proof of endpoint move invariance ${\tiny \square}$

\item {\em Invariance under vertex translations.}

Here, we consider three dual face $f_i$, $i=1,2,3$, of $(\T^*, \partial \T^*)$ each containing $n_i$ boundary edges positively oriented from vertex $1$ to vertex $n_i$. The three faces meet on the common edge $e$ which is such that $1=t(e)$, i.e., $e = e^i_{n_i}$, forall $i$.
Suppose that there are $p_1 < n_1$ dual edges $e^1_{1},...,e^1_{p_1}$ (resp. $p_2 < n_2$ dual edges $e^2_{1},...,e^2_{p_2}$) of the face $f_1$ (resp. $f_2$) supporting a string spin network edge $e_{\Gamma}^1$ (resp. $e_{\Gamma}^2$) colored by a spin $j_1$ (resp. $j_2$) representation and oriented negatively w.r.t the orientation of $f_1$ (resp. $f_2$).
Suppose also that the face $f_3$ contains $n_3-p_3$ dual edges $e^3_{p_3+1},...,e$ along which lies a positively oriented (w.r.t. to the orientation of the face) string spin network edge $e_{\Gamma}^3$ colored by a spin $j_3$ representation.
Consider that the three edges meet on the vertex $v_{\Gamma}$ supported by the vertex $1$ of $(\T^*, \partial \T^*)$.
Noting $g$ the group element associated to the common edge $e$, one can write the spin network function associated to the three valent vertex $v_{\Gamma}$ lying on $1$ and use the invariance property of the associated intertwining operator $\iota$ to `slide' the vertex along the edge $e$ :
\beqa
&& \stackrel{j_1}{\pi} ((g^1_{p_1})^{-1} (g^1_{p_1-1})^{-1} ... (g^1_{1})^{-1}) \, \stackrel{j_2}{\pi} ((g^2_{p_2})^{-1} (g^2_{p_2-1})^{-1}... (g^2_{1})^{-1}) \\ \nn
&& \stackrel{j_3}{\pi} (g^3_{p_3+1} ... g^3_{n_3-1} g) \,\,\, \iota_{j_1j_2j_3} \\ \nn
&=& \stackrel{j_1}{\pi} ((g^1_{p_1})^{-1} ... (g^1_{1})^{-1} g^{-1}) \, \stackrel{j_2}{\pi} ((g^2_{p_2})^{-1} ... (g^2_{1})^{-1} g^{-1}) \,
\stackrel{j_3}{\pi} (g^3_{p_3+1} ... g^3_{n_3-1}) \,\,\, \iota_{j_1j_2j_3}.
\eeqa
It is then possible to use the flatness constraint $F=0$ on either of the faces $f_1$ or $f_2$ to implement an edge move on $e_{\Gamma}^1$ or $e_{\Gamma}^2$ thus completing the vertex move ${\tiny \square}$
\end{itemize}

By virtue of all the above derivations, we have now fully proven theorem $1$.

To be perfectly complete, we need to verify that the amplitudes are also independent under orientation and holonomy base point change. This leads to the following proposition.

\begin{Pn}
The regularized physical inner product \eqref{regulfinal2} is independent of the choice of orientations of the dual edges and faces of $(\T^*,\partial \T^*)$, and does not depend on the choice of holonomy base points.
\end{Pn}

{\itshape Proof of Proposition $1$.}
It is immediate to see that the amplitudes do not depend on the orientations of the dual faces and dual edges of $(\T^*\partial \T^*)$, nor on the holonomy starting points on the boundaries of the dual faces \cite{pr2}. Indeed, a dual face and dual edge orientation change correspond respectively to a change $g_e \mapsto g_e^{-1}$ and $g_f \mapsto g_f^{-1}$ which are respectively compensated by the invariance of the Haar measure, $d g_e = dg_e^{-1}$, and of the delta function: $\delta(g_f) = \delta(g_f^{-1})$. A change in the holonomy base point associated to a dual face $f$ will have as a consequence the conjugation of the group element $g_f$ by some element $h$ in $G$. Since the delta function is central, $\delta(g_f) = \delta(h g_f h^{-1})$, the regularized physical inner product \eqref{regulfinal2} will remain unchanged under a such transformation.

Concerning the base point $x$ used to define the holonomies along the loop $\alpha$ and the paths $\beta \in C$ in \eqref{regulfinal2}, the situation is similar. Let $x$ be noted $x_1$ and suppose that we change the point $x_1$ to another point $x_2$ in $X$ neighbouring $x_1$. Since we have showen the invariance under bistellars and shellings, we are free to choose the simplest discretization \footnote{Anticipating on the next paragraph, we are using the fact that the invariance under Pachner moves implies the topological invariance of the amplitudes. Accordingly, we can use a cellular decomposition which is not necessarily a triangulation.} of the manifold $(\M,T_{\eta})$. We choose it such that the cylindrical section of $T_{\eta}$ between $x_1$ and $x_2$ is discretized by a single dual face with two opposite sides glued along a dual edge $e$ linking $x_1$ to $x_2$. By virtue of the curve move invariance, we are also free to choose the path $\beta$ to be along $e$. The amplitude based on $x_2$ as a starting point for the paths $\alpha$ and $\beta$, restricted to this section of $T_{\eta}$, yields
\beq
\int_{G^5} \prod_{a=1}^2 dg_a d\lambda_a d g_{\beta} \, \delta(g_2 \lambda_2 u \lambda_2^{-1}) \, \delta(g_{\beta} \lambda_1 \lambda_2^{-1}) \, \delta(g_2 g_{\beta} g_1^{-1} g_{\beta}^{-1}) \, f(\{g_a\}_a,\{\lambda_a\}_a,g_{\beta}),
\eeq
where $g_a$ is the holonomy around the disk bounding the tube section at the point $x_a$ and the function $f$ describes the string spin network function together with the other delta functions containing the group elements $g_a$ and $g_{\beta}$.
It is immediate to rewrite the above expression as
\beq
\int_{G^5} \prod_{a=1}^2 dg_a d\lambda_a d g_{\beta} \, \delta(g_1 \lambda_1 u \lambda_1^{-1}) \, \delta(g_{\beta}^{-1} \lambda_2 \lambda_1^{-1}) \, \delta(g_1^{-1} g_{\beta}^{-1} g_2 g_{\beta}) \, f(\{g_a\}_a,\{\lambda_a\}_a,g_{\beta}),
\eeq
which is the amplitude based on $x_1$ as a starting point for the paths $\alpha$ and $\beta$ {\small $\square$}

\vspace{2mm}

There are two major consequences due to the above theorem and
proposition. First, there is no continuum limit to be taken in
\eqref{regulfinal1}. Since the transition amplitudes are invariant
under elementary regulator moves, the regularized physical inner
product \eqref{regulfinal2} is independent of the regulator and
the expression \eqref{regulfinal2} is consequently exact, there is
no need to take the limits \footnote{More precisely, we have shown
that \eqref{regulfinal2} is invariant when going from $\Reg$ to
$\mathcal{R}_{(\eta',\epsilon')}$ for all $(\eta', \epsilon') \neq
(\eta, \epsilon)$ which implies regulator
independence.}  $\epsilon, \eta \rightarrow 0$. In particular, we
have shown that the amplitudes are invariant under any finite
sequence of bistellar moves and shellings which implies, by
Pachner's theorem, that the physical inner product is well defined
and invariant on the equivalence classes of PL-manifolds
\footnote{More precisely, a triangulation $\Delta$ is not a
PL-manifold. It is a combinatorial manifold which is PL-isomorphic
to a PL-manifold.} $(\T,\partial \T)$ up to PL-homeomorphisms.
Accordingly, the transition amplitudes are invariant under
triangulation change and thus under refinement. This leads to the
second substantial consequence of theorem $1$. The crucial point
is that the equivalence classes of PL-manifolds up to
PL-homeomorphism are in one-to-one correspondence with those of
topological manifolds up to homeomorphism. See for instance
\cite{pfeiffer} for details. Hence, showing the invariance of the
regularized physical inner product under triangulation change is
equivalent to showing homeomorphism invariance: the discretized
expression \eqref{regulfinal2} is in fact a topological invariant
of the manifold $(\M,T_{\eta})$. In particular, the amplitudes are
invariant on the equivalence classes of boundary torii $T_{\eta}$
up to homeomorphisms. It follows that they do not depend on the
embedding of the string $\Ss$.

Combining these results with the second part of theorem $1$ stating that the regularized physical inner product is invariant under string spin netwotk graph moves, we obtain the following corollary.

\begin{co}
The physical inner product \eqref{regulfinal2} is a topological invariant of the triple $((\M,T_{\eta}),\Gamma)$ :
\beq
P[\Reg;\Psi_{\Gamma}] = P[[(\M,T_{\eta})];\Psi_{[\Gamma]}],
\eeq
\\
where $[(\M,T_{\eta})]$ and $[\Gamma]$ denote the equivalence classes of topological open manifolds and one-complexes up to homeomorphisms and ambient isotopy respectively.
\end{co}

This corollary concludes our study of the topological invariance of the theory of extended matter coupled to BF theory studied in this paper.

\section{Conclusion}

In the first part of this paper we have studied the geometrical
interpretation of the solutions of the BF theory with string-like
conical defects. We showed the link between solutions of our
theory and solutions of general relativity of the cosmic string
type. We provided a complete geometrical interpretation of the
classical string solutions and explained (by analyzing the multiple
strings solution) how the presence of strings at different
locations induces torsion. In turn torsion can in principle be
used to define localization in the theory.

We have achieved the full background independent quantization of
the theory introduced in \cite{BP}. We showed that the
implementation of the dynamical constraints at the quantum level
require the introduction of regulators. These regulators are
defined by (suitable but otherwise arbitrary) space
discretization. Physical amplitudes are independent of the
ambiguities associated to the way this regulator is introduced and
are hence well defined. There are other regularization ambiguities
arising in the quantization process that have not been explicitly
treated here. For an account of these as well as for a proof that
these have no effect on physical amplitudes see
\cite{Perez:2005fn}.

The results of this work can be applied to the more general type
of models introduced in \cite{merced}, were it is shown that a
variety of physically interesting 2-dimensional field theories can
be coupled to the string world sheet in an consistent manner. An
interesting example is the one where in addition to the degrees of
freedom described here, the world sheet carries Yang-Mills
excitations.

There is an intriguing connection between this type of topological
theories and certain field theories in the 2+1 gravity plus
particles case. One would expect a similar connection to exist in
this case. However, due to the higher dimensional character of the
excitations in this model this relationship allows for the
inclusion of more general structures: only spin and mass is
allowed in 2+1 dimensions. The study of the case involving
Yang-Mills world-sheet degrees of freedom is of special interest.
This work provides the basis for the computation of amplitudes in
the topological theory. A clear understanding of the properties of
string transition amplitudes should shed light on the eventual
relation with field theories with infinitely many degrees of
freedom.

\subsection*{Acknowledgements}
We thank Romain Brasselet for his active participation to the early stages of this project.
WF thanks Etera Livine for discussions. This work was supported in part by the 
Coordena\c{c}\~{a}o de Aperfei\c{c}oamento de Pessoal de 
N\'{\i}vel Superior (Capes) through a visiting professor fellowship.

\end{document}